\documentclass{book}
\usepackage{makeidx,epsfig}
\usepackage{setspace,graphicx}%,subfigure
\usepackage{Generic}
\makeindex

\newcommand{\rd}{{\rm d}}

\newcommand{\ri}{{\rm i}}
\newcommand{\kc}{k_{\rm c}}
\newcommand{\kL}{k_{\rm L}}
\newcommand{\Er}{E_{\rm rec}}
\begin{document}
\title{Dynamic localization in optical lattices\footnote{Chapter 12 of the
	volume: {\bf Dynamical Tunneling - Theory and Experiment}, 
	\newline \hspace*{1.6em}
	edited by S.~Keshavamurthy and P.~Schlagheck 
	(Taylor and Francis CRC, 2011)
	\newline $~$}}
\author{Stephan Arlinghaus, 
	Matthias Langemeyer, 
	and Martin Holthaus\footnote{Carl von Ossietzky Universit\"at
	\newline \hspace*{1.6em}
	Institut f\"ur Physik
	\newline \hspace*{1.6em}
	D - 26111 Oldenburg, Germany}}

\pagenumbering{roman}
\begin{doublespace}
\maketitle 
\frontmatter
\tableofcontents
\mainmatter
\pagenumbering{arabic}
\chapter{Dynamic localization in optical lattices}

The concept of dynamic localization goes back to an observation reported
by Dunlap and Kenkre in 1986: The wave packet of a single particle moving 
on a single-band tight-binding lattice endowed with only nearest-neighbor 
couplings remains perpetually localized when driven by a spatially homogeneous
ac force, provided the amplitude and the frequency of that force obey a 
certain condition~\cite{DunlapKenkre86}. When trying to overcome the 
limitations of the model, it is comparatively straightforward to deal 
with an arbitrary form of the dispersion relation --- thus abandoning the 
nearest-neighbor approximation --- and with arbitrary time-periodic forces, 
thus doing away with the restriction to purely sinusoidal 
driving~\cite{DignamSterke02}. But in any real lattice system an external
time-periodic force will induce interband transitions, and it is by no means
obvious whether dynamic localization can survive when these come into play.

In this chapter we consider ultracold atoms in driven optical lattices,
which provide particularly attractive, experimentally well accessible 
examples of quantum particles in spatially periodic structures exposed to 
time-periodic forcing~\cite{LignierEtAl07,EckardtEtAl09,ZenesiniEtAl09}.
Such systems are much cleaner, and more easy to control, than electrons
in crystal lattices under the influence of ac electric fields, for which
the original idea had been developed~\cite{DunlapKenkre86}. With the help 
of results obtained by numerical calculations we illustrate that such 
ultracold atoms in kHz-driven optical lattices exhibit dynamic localization
in almost its purest form if the parameters are chosen judiciously, despite
the potentially devastating presence of interband transitions.  

When viewing dynamic localization as resulting from a band 
collapse~\cite{Holthaus92,HolthausHone93}, far-reaching further 
possibilities emerge. Namely, the actual strength of deviations from 
exact spatial periodicity, be they isolated~\cite{HoneHolthaus93}, 
random~\cite{HolthausEtAl95}, or 
quasiperiodic~\cite{DreseHolthaus97a,DreseHolthaus97b}, is measured relative 
to the effective band width. Thus, when the band in question almost collapses 
in response to time-periodic driving, the effects of even slight deviations 
from exact lattice periodicity are strongly enhanced. This allows one, 
in particular, to coherently control the ``metal-insulator''-like
incommensurability transition occurring in sufficiently deep quasiperiodic 
optical lattices~\cite{DreseHolthaus97a,DreseHolthaus97b,BoersEtAl07}. While 
the very transition has already been observed with Bose-Einstein condensates 
in bichromatic optical potentials~\cite{RoatiEtAl08}, its coherent control by 
means of time-periodic forcing still awaits its experimental verification.

\section{The basic idea}

The one-dimensional tight-binding system described by the Hamiltonian
\begin{equation}
	H_0 = -J \sum_{\ell} \Big( |\ell + 1 \rangle \langle \ell |
	+ | \ell \rangle \langle \ell + 1 | \Big) \; ,
\label{eq:TBS}
\end{equation}
where $|\ell\rangle$ denotes a Wannier state localized at the $\ell$th
lattice site, and $J$ is the hopping matrix element connecting neighboring
sites, is about the simplest model for the formation of Bloch bands.
Assuming that the unspecified number of sites is so large that finite-size
effects may be neglected, its energy eigenstates are Bloch waves 
\begin{equation}
	| \varphi_k \rangle = \sum_\ell | \ell \rangle \exp(\ri \ell k a) 
\end{equation}
labeled by a wave number $k$; the lattice period is given by $a$. The
corresponding energy dispersion relation reads
\begin{equation}
	E(k) = -2J \cos(ka) \; ;
\label{eq:EDR}	
\end{equation}
here we assume $J > 0$, so that its minimum is located at $k = 0 \bmod 2\pi/a$.
Now we let an external time-dependent, spatially homogeneous force $F(t)$ act 
on the system, such that the total Hamiltonian becomes 
\begin{equation}
	H(t) = H_0 + H_1(t)
\label{eq:HFL}	
\end{equation}
with
\begin{equation}
	H_1(t) = - F(t) \sum_\ell | \ell \rangle a \ell \langle \ell | \; .
\label{eq:HIN}
\end{equation}
It is easy to verify that the wave functions 
\begin{equation}
	| \psi_k(t) \rangle = 
	\exp\!\left( -\frac{\ri}{\hbar} \int_0^t \! \rd \tau \, 
	E\big( q_k(\tau) \big) \right)
	\sum_\ell | \ell \rangle \exp\big(\ri \ell q_k(t) a \big) 
\label{eq:GHS}
\end{equation}	
then are solutions to the time-dependent Schr\"odinger equation, provided 
the time-dependent wave numbers $q_k(t)$ introduced here obey the 
``semiclassical'' relation 		
\begin{equation}
	\hbar \dot{q}_k(t) = F(t) \; .
\label{eq:IXH}	
\end{equation}
We demand that $q_k(t)$ be equal to $k$ at time $t = 0$, and therefore
set
\begin{equation}		
	q_k(t) = k + \frac{1}{\hbar} \int_0^t \! \rd \tau \, F(\tau) \; .
\end{equation}	
These wave functions~(\ref{eq:GHS}), originally considered by Houston
in the context of crystal electrons exposed to a uniform electric
field superimposed on a periodic lattice potential~\cite{Houston40},
are known as ``accelerated Bloch waves'', or Houston states. 

In the particular case of a monochromatic force with angular 
frequency~$\omega$ and amplitude~$F_1$, given by 
\begin{equation}
	F(t) = F_1 \cos(\omega t) \; ,
\label{eq:TPF}	
\end{equation}	
one has
\begin{equation}
	q_k(t) = k + \frac{F_1}{\hbar\omega} \sin(\omega t) \; , 
\end{equation}	
so that $q_k(t)$ naturally acquires the temporal period $T = 2\pi/\omega$
of the driving force. Then also $E\big( q_k(t) \big)$ is $T$-periodic,
but the Houston state~(\ref{eq:GHS}) is not, because the integral appearing 
in the exponential prefactor acquires a contribution which grows linearly 
with time. In order to extract that contribution, we calculate the one-cycle
average 
\begin{eqnarray}
	\varepsilon(k) & \equiv & \frac{1}{T} \int_0^T \! \rd \tau \,
	 E\big( q_k(\tau) \big)
\nonumber	\\     & = & 
	-2 J_{\rm eff} \cos(ka) \; ,
\label{eq:QED}
\end{eqnarray}
thus obtaining an effective hopping matrix element given by
\begin{equation}
	J_{\rm eff} = 
	J \, {\rm J}_0\! \left( \frac{F_1 a}{\hbar\omega} \right) \; ,
\label{eq:MHE}
\end{equation}	
with ${\rm J}_0(z)$ denoting the Bessel function of zero order. We then
write 
\begin{equation}
	\exp\!\left( -\frac{\ri}{\hbar} \int_0^t \! \rd \tau \, 
	E\big( q_k(\tau) \big) \right) = 
	\exp\!\left( -\frac{\ri}{\hbar} \int_0^t \! \rd \tau \,
	\Big[ E\big( q_k(\tau) \big) - \varepsilon(k) \Big] \right)
	\exp\!\Big(-\ri \varepsilon(k) t / \hbar\Big) \; ,
\label{eq:DCS}
\end{equation}
so that the first exponential on the right hand side now is $T$-periodic 
by construction. Hence, for the $T$-periodic force~(\ref{eq:TPF}) the 
Houston states~(\ref{eq:GHS}) can be cast into a form 
\begin{equation}
	| \psi_k(t) \rangle = | u_k(t) \rangle 
	\exp\!\Big(-\ri \varepsilon(k) t / \hbar\Big)
\label{eq:FHS}
\end{equation}	
with $T$-periodic functions $| u_k(t) \rangle$,   
\begin{equation}
	| u_k(t) \rangle = | u_k(t+T) \rangle \; . 
\end{equation}	
This leads to a remarkable conclusion. Any wave packet governed by the full
Hamiltonian~(\ref{eq:HFL}) with periodic forcing~(\ref{eq:TPF}) can be 
expanded with respect to these states~(\ref{eq:FHS}) with coefficents that 
are constant in time, because the time-dependence already is fully 
incorporated into the states themselves. After each cycle~$T$ the $T$-periodic
functions $|u_k(t)\rangle$ are restored, so that the time evolution of the
wave packet, when viewed stroboscopically at intervals~$T$, is determined
by the different ``speed'' of rotation of the complex phase factors
$\exp(-\ri \varepsilon(k) t / \hbar)$ of the packet's individual components. 
But if all quantities $\varepsilon(k)$ are equal, which according to 
Eqs.~(\ref{eq:QED}) and (\ref{eq:MHE}) occurs when the scaled driving 
amplitude
\begin{equation}
	K_0 \equiv \frac{F_1 a}{\hbar\omega}
\label{eq:SCA}	
\end{equation} 
equals a zero of the Bessel function ${\rm J}_0$, all phase factors evolve
at the same speed, so that the wave packet reproduces itself exactly after
each period: There is some $T$-periodic wiggling, but no long-term motion.
This, in short, is dynamic localization~\cite{DunlapKenkre86}.

The above argument appears so special, and the decisive step~(\ref{eq:DCS})
so swift, that it is not easy to see how to transfer this finding to more 
realistic situations: How can one incorporate deviations from exact lattice 
periodicity into this reasoning? How to proceed when several bands are coupled 
by interband transitions? The answer to these questions is provided by the 
Floquet picture, which does not directly take recourse to the spatial lattice 
periodicity, but rather builds on the temporal periodicity of the Hamiltonian:
When $H(t) = H(t+T)$, there exists a complete set of solutions to the 
time-dependent Schr\"odinger equation of the particular form
\begin{equation}
	|\psi_n(t)\rangle = | u_n(t) \rangle 
	\exp(-\ri \varepsilon_n t/\hbar) \; ,
\label{eq:FLS}
\end{equation} 
where the functions $|u_n(t)\rangle = |u_n(t+T)\rangle$ inherit the 
$T$-periodicity of the underlying Hamiltonian. These states are known 
as Floquet states; the quantities $\varepsilon_n$ are dubbed as
quasienergies~\cite{Shirley65,Zeldovich67,Ritus67,Sambe73}. Obviously
the Houston states~(\ref{eq:GHS}) with time-periodic forcing are particular
examples of such Floquet states; from now on we employ an abstract state 
label~$n$ instead of the wave number~$k$ in order to also admit settings 
without lattice periodicity. In the case of the Houston-Floquet states, 
the determination of their quasienergies~(\ref{eq:QED}) essentially was a 
by-product of the solution of an initial value problem. The general case, 
however, has to proceed along a more sophisticated route: Floquet states 
and quasienergies are determined by solving the eigenvalue problem
\begin{equation}
	\left( H(t) - \ri\hbar\frac{\partial}{\partial t} \right) 
	| u_n (t) \rangle \! \rangle
	= \varepsilon_n | u_n (t) \rangle \! \rangle \; ,
\label{eq:EVP}
\end{equation}	   
posed in an {\em extended Hilbert space\/} of $T$-periodic functions;
in that space time plays the role of a {\em coordinate\/}. Therefore, if
$\langle u_1(t) | u_2(t) \rangle$ is the scalar product of two $T$-periodic
functions in the usual physical Hilbert space, their scalar product in
the extended space reads~\cite{Sambe73}
\begin{equation}
	\langle \! \langle u_1 | u_2 \rangle \! \rangle \equiv 
	\frac{1}{T} \int_0^T \! \rd t \, \langle u_1(t) | u_2(t) \rangle \; .
\label{eq:SCP}	
\end{equation}
Hence, we write $|u_n(t)\rangle$ for a Floquet eigenfunction when viewed in
the physical Hilbert space, and $|u_n(t)\rangle\!\rangle$ when that same
function is regarded as an element of the extended space.

A most important consequence of this formalism stems from the fact that when 
$| u_n (t) \rangle\!\rangle$ is a solution to the problem~(\ref{eq:EVP}) 
with eigenvalue $\varepsilon_n$, then  
$| u_n (t) \exp(\ri m \omega t) \rangle\!\rangle$ is a further solution
with eigenvalue $\varepsilon_n + m \hbar\omega$,  where we have set 
$\omega = 2\pi/T$, and $m$ is any (positive, zero, or negative) integer, 
in order to comply with the required $T$-periodic boundary condition. For 
$m \neq 0$ these two solutions are orthogonal with respect to the scalar 
product~(\ref{eq:SCP}). But when going back to the physical Hilbert space, 
one has   
\begin{equation} 
	| u_n(t) \exp(\ri m \omega t) \rangle
	\exp\!\Big(-\ri(\varepsilon_n + m\hbar\omega)/\hbar\Big) =  
	| u_n(t) \rangle \exp(-\ri \varepsilon_n t/\hbar) \; ,  	
\end{equation}
so that the two different solutions represent {\em the same\/} Floquet 
state~(\ref{eq:FLS}). We conclude that a physical Floquet state does 
not simply correspond to an individual solution to the eigenvalue 
problem~(\ref{eq:EVP}), but rather to a whole class of such solutions
labeled by the state index~$n$, whereas the ``photon'' index~$m$
distinguishes different representatives of such a class. Likewise, a 
quasienergy should not be regarded as a single eigenvalue, but rather as 
a set $\{\varepsilon_n + m\hbar\omega \; | \; m = 0,\pm 1,\pm2,\ldots\}$ 
associated with one particular state~$n$, while $m$ ranges through all 
integers. Therefore, each ``quasienergy Brillouin zone'' of width 
$\hbar\omega$ contains one quasienergy representative of each state.  

The time evolution of any wave function can then be written as a
Floquet-state expansion,
\begin{equation}
	|\psi(t)\rangle = \sum_n c_n | u_n(t) \rangle
	\exp(-\ri \varepsilon_n t/\hbar) \; ,      
\end{equation}
where the coefficients $c_n$ remain constant in time. This is one of the
main benefits offered by the Floquet picture, and allows one to draw
many parallels to the evolution of systems governed by a time-independent
Hamiltonian.

Equipped with this set of tools, it is now clear how to investigate the
possible occurrence of dynamic localization in realistic lattice structures:
One has to solve the eigenvalue problem~(\ref{eq:EVP}) for the Hamiltonian
with the respective full lattice potential, and to enquire whether the 
resulting quasienergy bands collapse at least approximately, that is, acquire 
negligible widths for certain parameters. If so, any wave packet prepared in 
a quasienergy band at a collapse point will suffer from ``prohibited
dephasing'', as in the archetypal model specified by Eqs.~(\ref{eq:TBS}),
(\ref{eq:HIN}), and (\ref{eq:TPF}); and thus remain dynamically localized. 
Interband transitions then are automatically included, with multiphoton-like 
resonances manifesting themselves through quasienergy-curve 
anticrossings~\cite{ArlinghausHolthaus10}.

In the following section we will carry through this program for ultracold
atoms in driven one-dimensional optical lattices.

\section{Does it work?}

A one-dimensional optical lattice is created by two counterpropagating laser 
beams with wave number $k_{\rm L}$, suitably detuned from a dipole-allowed 
transition of the atomic species moving in this standing light wave. By means 
of the ac Stark effect, the spatially periodic electric field experienced by
the atoms then translates into a cosine potential
\begin{equation}
	V_{\rm lat}(x) = \frac{V_0}{2}\cos(2k_{\rm L} x)
\label{eq:VOL}	 
\end{equation} 
for their translational motion along the lattice, with a depth~$V_0$ that 
is proportional to the laser intensity~\cite{MorschOberthaler06,BlochEtAl08}. 
The characteristic energy scale then is given by the single-photon recoil 
energy
\begin{equation}
	\Er = \frac{\hbar^2 \kL^2}{2M} \; ,	
\end{equation}   
where $M$ denotes the atomic mass. To give a numerical example, when working 
with $^{87}$Rb in a lattice generated by laser radiation with wavelength 
$\lambda = 2\pi/\kL = 842$~nm~\cite{EckardtEtAl09,ZenesiniEtAl09} one has 
$\Er = 1.34 \cdot 10^{-11}$~eV. Thus, typical lattice depths of 5 to 10
recoil energies are on the order of $10^{-10}$~eV --- which means that one 
encounters many phenomena with ultracold atoms in optical lattices which are 
known from traditional solid-state physics, but scaled down in energy by no 
less than 10~orders of magnitude. 

This also tells us what ``ultracold'' means. Taking an ensemble of atoms with 
a temperature~$T_{\rm ens}$ such that $k_{\rm B} T_{\rm ens}$ is roughly equal 
to $\Er$, say, where $k_{\rm B}$ is Boltzmann's constant, the de Broglie 
wavelength of these atoms, given by
\begin{equation}
	\lambda_{\rm de Broglie} 
	= \frac{h}{\sqrt{2\pi M k_{\rm B} T_{\rm ens}}}
	\approx \frac{2}{\sqrt{\pi}} \, \frac{\lambda}{2} \; , 
\end{equation}
is barely longer than the lattice constant $a = \lambda/2$. But in order to 
experience quantum mechanical lattice effects, the particles have to be able
to ``feel'' the periodic structure, so that $\lambda_{\rm de Broglie}$ 
should cover {\em at least\/} a few lattice constants --- which means that 
being this cold is not cold enough: We even require 
$k_{\rm B} T_{\rm ens} \ll \Er$.

With hardly any thermal excitation energy left the atoms occupy only the 
lowest Bloch band of their optical lattice, so that the single-particle 
Hamiltonian with the lattice potential~(\ref{eq:VOL}) translates directly 
into the single-band tight-binding model~(\ref{eq:TBS}) when working in a 
basis of Wannier functions pertaining to that lowest band, and neglecting 
all couplings other than those between nearest neighbors, denoted as~$J$.
The accuracy of this approximation increases with increasing lattice 
depth~\cite{EckardtEtAl09,BoersEtAl07}: For $V_0/\Er = 5$ the magnitude of 
the ratio of the neglected matrix element connecting next-to-nearest neighbors 
to $J$ still reaches about 5\%, but it decreases to about 1\% when 
$V_0/\Er = 10$. Moreover, when expressing the exact band structure of a 
cosine lattice in terms of characteristic values of the Mathieu equation, 
and noting that the width~$W$ of the cosine energy band~(\ref{eq:EDR}) is 
$4J$, one finds the approximation~\cite{BlochEtAl08}
\begin{equation}
	J/\Er \sim \frac{4}{\sqrt{\pi}}\left(\frac{V_0}{\Er}\right)^{3/4}
	\exp\left(-2\sqrt{\frac{V_0}{\Er}}\right)
	\qquad \mbox{for} \quad V_0/\Er \gg 1 \; .
\label{eq:MAP}
\end{equation}  
The requisite still missing now is the time-periodic force corresponding to the
model~(\ref{eq:HIN}). This can be effectuated either by introducing a small 
oscillating frequency difference between the two lattice-generating laser beams,
as detailed later, or by retro-reflecting one such beam off an oscillating 
mirror back into itself~\cite{LignierEtAl07,EckardtEtAl09,ZenesiniEtAl09}. 
In a frame of reference co-moving with the oscillating lattice, one then 
obtains the single-particle Hamiltonian
\begin{equation}
	H(t) = \frac{p^2}{2M} + V_{\rm lat}(x)
	 - F_1 x \cos(\omega t + \phi) \; ,
\label{eq:HDL}	
\end{equation}
where $p$ is the atomic center-of-mass momentum in the lattice direction,
the driving force is parametrized in accordance with Eq.~(\ref{eq:TPF}),
and we have also admitted an arbitrary phase~$\phi$.

\begin{figure}[t!]
\centering
\includegraphics[width=1.0\textwidth]{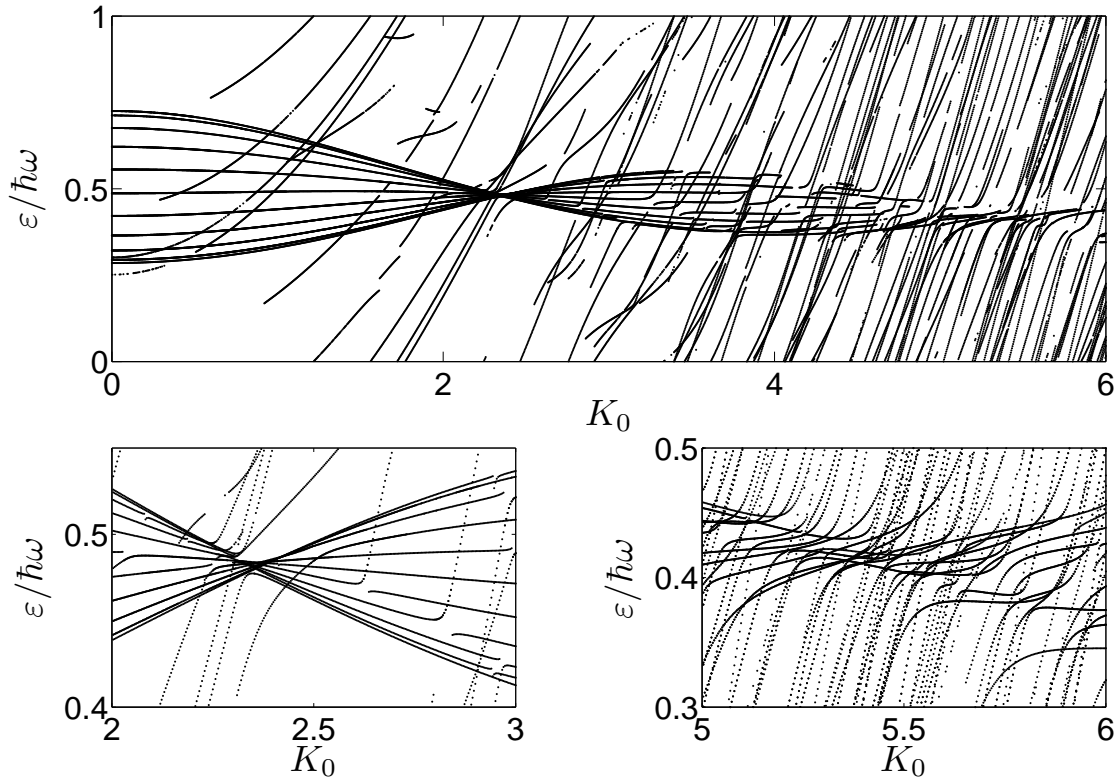}
\caption{Above: One Brillouin zone of quasienergies for the optical 
	lattice~(\ref{eq:VOL}) with depth $V_0/\Er = 5.7$, driven with
	scaled frequency $\hbar\omega/\Er = 0.5$, vs.\ the scaled driving
	amplitude~$K_0$. The lower left panel testifies that the first band 
	collapse is almost perfect, whereas the second one, enlarged in the 
	lower right panel, is already thwarted by multiphoton-like resonances.}  
\label{fig:F_1}
\end{figure}

In all our model calculations we consider a lattice with depth 
$V_0/\Er = 5.7$, implying that the width of the lowest Bloch band is 
$W/\Er = 0.22$, whereas the gap between this lowest band and the first 
excited one figures as $\Delta/\Er = 2.76$. Even for such a comparatively 
shallow lattice, which is routinely being realized in current 
experiments~\cite{ZenesiniEtAl09}, the dispersion of the lowest band 
already is reasonably well described by the tight-binding cosine 
approximation~(\ref{eq:EDR}), setting $J = W/4$. In order to obtain 
dynamic localization, the driving frequency should then be chosen such that 
the quantum $\hbar\omega$ is significantly smaller than the gap~$\Delta$, 
so that, perturbatively speaking, interband transitions require higher-order 
multiphoton-like processes, which would be suppressed as long as the driving 
amplitude $F_1$ is not too strong~\cite{ArlinghausHolthaus10}. On the other 
hand, it is reasonable to demand that $\hbar\omega$ be larger than the band 
width, so that the band fits into a single quasienergy Brillouin zone. A good 
choice of the driving frequency should therefore adhere to the chain 
$4J = W < \hbar\omega < \Delta$; we take $\hbar\omega/\Er = 0.5$ 
in all numerical scenarios depicted below. For $^{87}$Rb atoms in a lattice 
with $\lambda = 842$~nm this choice fixes the frequency at 
$\omega/(2\pi) = 1.62$~kHz.   
 
Figure~\ref{fig:F_1} shows one Brillouin zone of quasienergies for these
parameters vs.\ the scaled driving amplitude $K_0$, as defined by 
Eq.~(\ref{eq:SCA}). Observe that the first quasimomentum Brillouin zone ranges 
from $-\hbar\pi/a = -\hbar\kL$ to $+\hbar\pi/a = +\hbar\kL$; the homogeneous 
force does not mix states with different wave numbers~\cite{DreseHolthaus97a}. 
Hence, we combine  quasienergies for states with $k = (i/10) \, \kL$ in this 
plot, with $i = 0,1,2,\ldots,10$. In this way, the comparison of the ideal
quasienergy band~(\ref{eq:QED}) with the one appearing in the actual optical
lattice is greatly facilitated. Evidently the first band collapse is almost 
perfect, although it is slightly shifted from $K_0 = 2.405$, the first zero 
of ${\rm J}_0$, to $K_0 \approx 2.35$. In contrast, the second collapse, 
expected at $K_0 = 5.520$, already is significantly affected by a host 
of anticrossings, indicating multiphoton-like resonances. Thus, with 
$V_0/\Er = 5.7$ and $\hbar\omega/\Er = 0.5$ we may expect almost perfect
dynamic localization at the first collapse point, whereas there will be
strong disturbances of the ideal dynamics at the second one.

\begin{figure}[t!]
\centering
\includegraphics[width=1.0\textwidth]{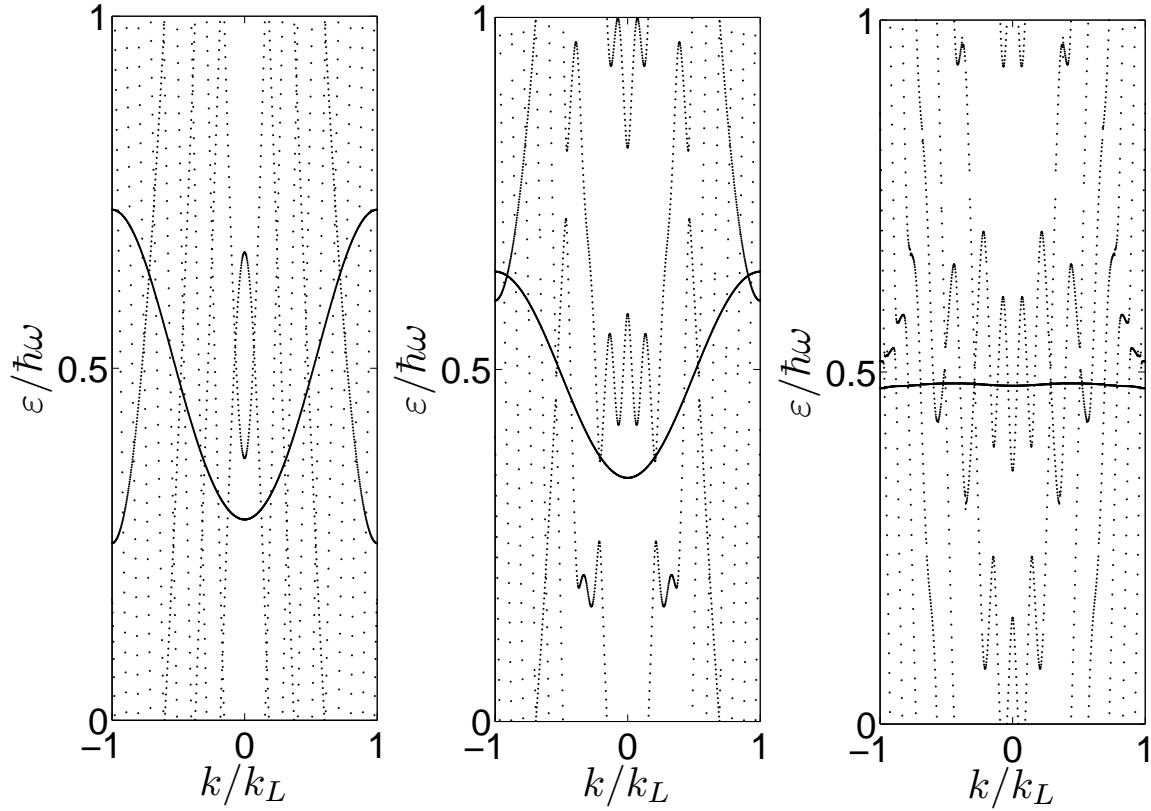}
\caption{``Lowest'' quasienergy band for the optical lattice~(\ref{eq:VOL}) 
	with depth $V_0/\Er = 5.7$, driven with scaled frequency 
	$\hbar\omega/\Er = 0.5$, and scaled amplitudes $K_0 = 0$ (left),
	$1.18$ (middle), and $2.35$ (right). Additional curves result
	from higher bands.}
\label{fig:F_2}
\end{figure}

In Fig.~\ref{fig:F_2} we depict the lowest quasienergy band for $K_0 = 0$,
where it coincides with the original energy band; $K_0 = 1.18$, where its
width is reduced by a factor of ${\rm J}_0(1.18) = 0.681$; and at the first 
collapse point, $K_0 = 2.35$. Ideally, a collapsed quasienergy band is 
completely flat, so that dynamic localization is associated with an infinite 
effective mass of the driven Bloch particle. Here we still observe some 
residual dispersion, probably resulting from both next-to-nearest neighbor
couplings and couplings to higher bands, but the degree of band flattening 
achieved by the driving force is nonetheless impressive.   

The ultimate demonstration of dynamic localization requires, of course, 
the inspection of wave-packet dynamics. To this end, we first compute the
Bloch states $\langle x | \varphi_{1,k} \rangle$ of the lowest energy band 
of the lattice~(\ref{eq:VOL}), and use them to design an initial wave packet
\begin{equation}
	\langle x | \psi(t\!=\!0) \rangle =
	\int_{-\kL}^{\kL} \! \rd k \, g_1(k,t\!=\!0) \, 
	\langle x | \varphi_{1,k} \rangle 
\label{eq:BWP}
\end{equation}
with a Gaussian $k$-space distribution
\begin{equation}	
	g_1(k,t\!=\!0) = \frac{1}{\sqrt{2 \kL \sqrt{\pi} \Delta k }}
	\exp\!\left(- \frac{(k - \kc)^2}{2 (\Delta k)^2} \right)
\label{eq:GKD}
\end{equation} 
centered around some predetermined wave number $\kc$, with width $\Delta k$.
The corresponding probability density $|\langle x | \psi(t\!=\!0) \rangle|^2$
is concentrated in the wells of the lattice potential, equipped with a 
Gaussian envelope that varies the more slowly with $x$ the narrower its 
distribution~(\ref{eq:GKD}), that is, the smaller $\Delta k$. We then 
take this packet~(\ref{eq:BWP}) as initial condition, and compute the 
wave function $\langle x | \psi(t) \rangle$ for $t > 0$ by solving the 
time-dependent Schr\"odinger equation numerically, fixing the phase $\phi$ 
in the Hamiltonian~(\ref{eq:HDL}) at the value $\phi = \pi/2$. This means that
the force $F(t) = F_1\cos(\omega t + \phi)$ is instantaneously switched on
at $t = 0$.

\begin{figure}[ht!]
\centering
\includegraphics[width=0.63\textwidth]{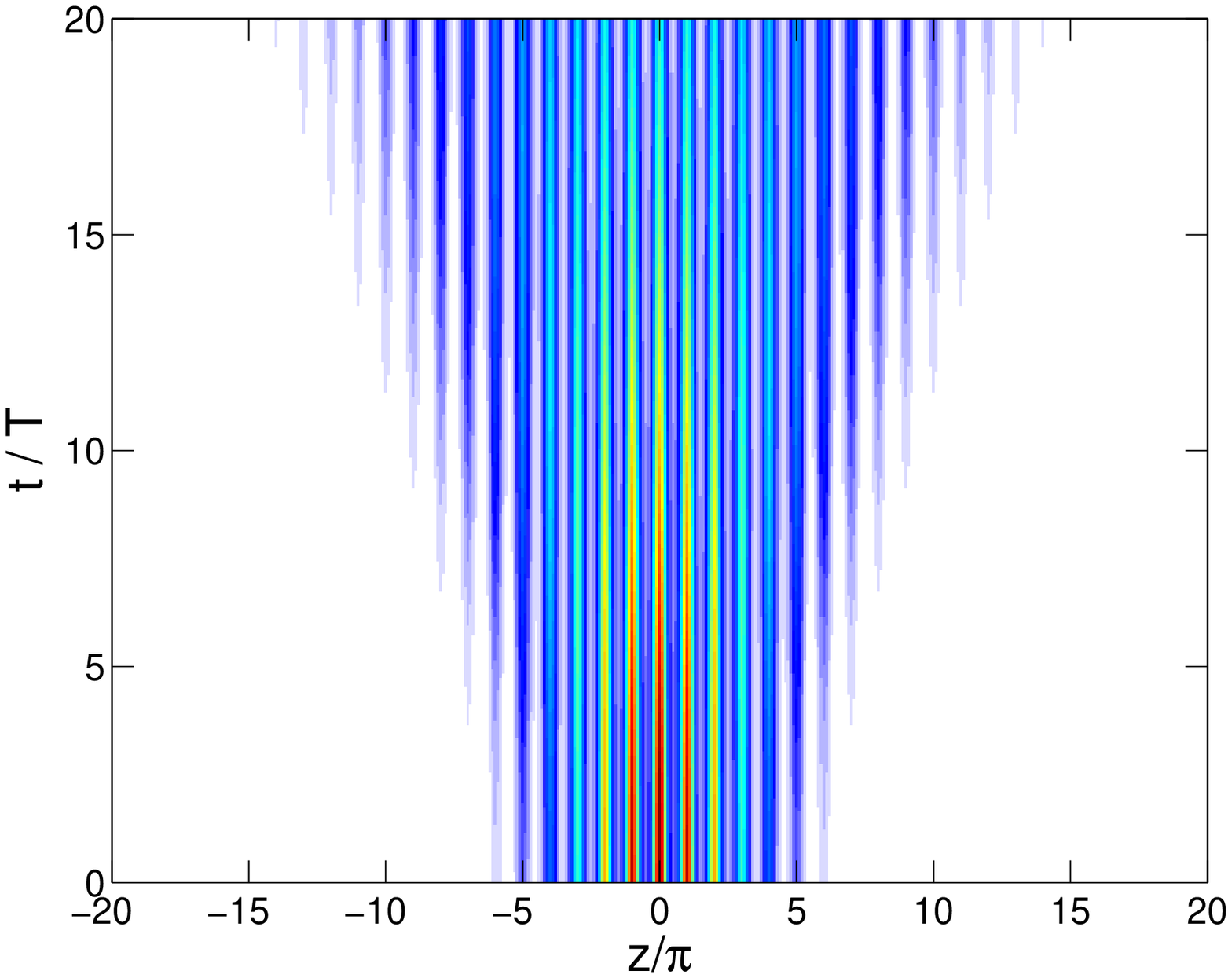}
\caption{Spreading of the Bloch wave packet~(\ref{eq:BWP}) with initial 
	$k$-space width $\Delta k/ \kL = 0.1$, and initial momentum 
	$\kc/\kL = 0$, in the unforced optical lattice. In this and the 
	following figures, density is encoded in shades of gray.} 
\label{fig:F_3}
%\end{figure}

\strut

%\begin{figure}[b]
\centering
\includegraphics[width=0.63\textwidth]{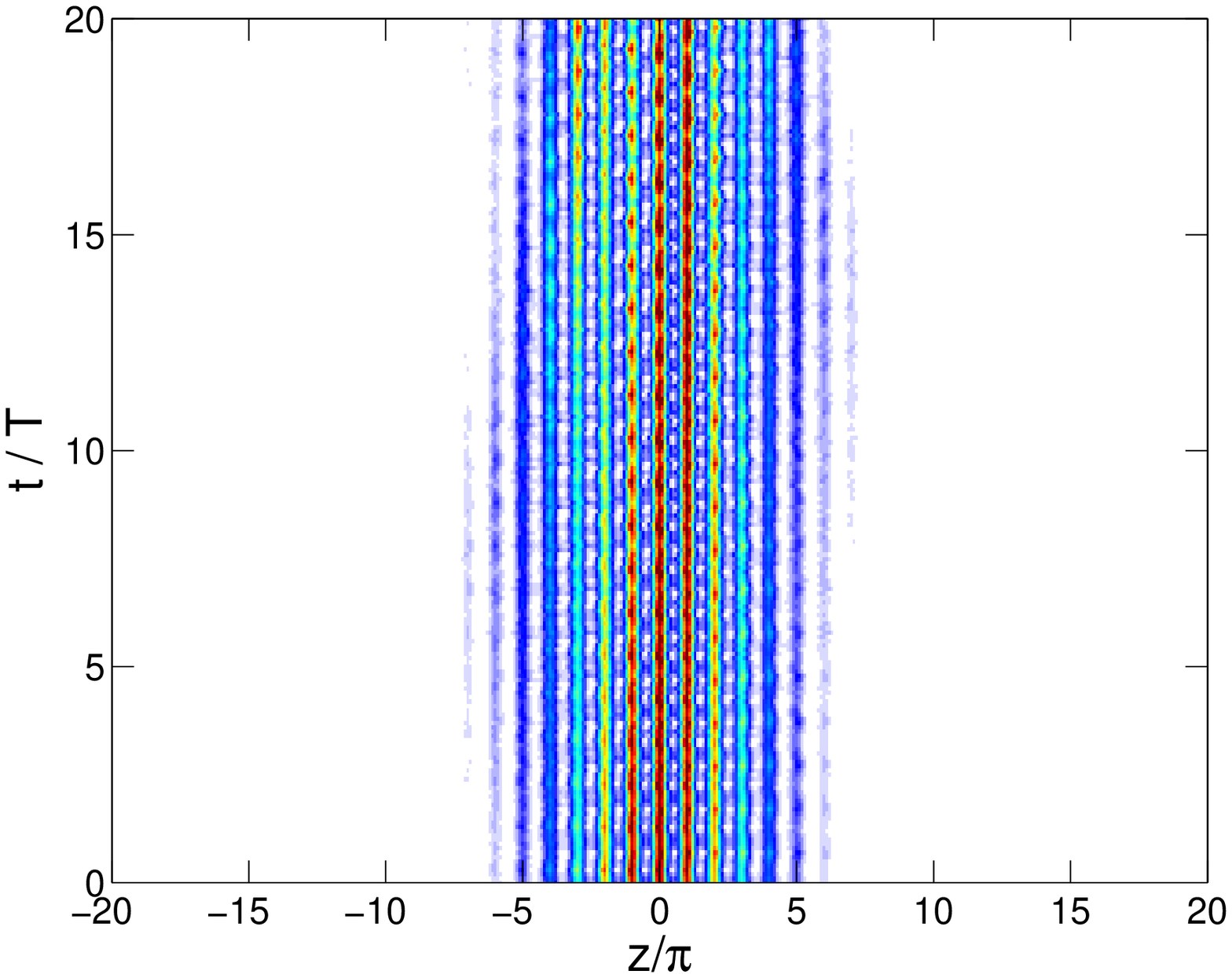}
\caption{Evolution of the same initial wave packet as in Fig.~\ref{fig:F_3}
	at the first band collapse ($K_0 = 2.35$): Here one encounters 
	almost perfect dynamic localization; wave-packet spreading is
	disabled because the quasienergy band is dispersionless.}  
\label{fig:F_4}
\end{figure}

Figure~\ref{fig:F_3} shows a density plot of the wave packet when it evolves 
in the undriven lattice, that is, for $K_0 = 0$; the density is encoded in 
shades of gray. In this and the following figures, spatial extensions are 
measured in terms of the dimensionless coordinate $z = \kL x$, so that a 
distance $\Delta z/\pi = 1$ corresponds to one lattice period; moreover, 
the time scale is set by the period $T = 2\pi/\omega$. With $\kc/\kL = 0$ 
the initial packet carries no net momentum; its width is chosen as 
$\Delta k/\kL = 0.1$. As expected, the width of the packet then grows in 
the course of time by well-to-well tunneling.

In Fig.~\ref{fig:F_4} we depict the density of the wave packet that evolves 
from the same initial condition when the driving amplitude is tuned to 
the first band collapse at $K_0 = 2.35$. Here we observe dynamic localization 
at its very best: The spreading has stopped, the packet is ``frozen''.

\begin{figure}[t!]
\centering
\includegraphics[width=0.63\textwidth]{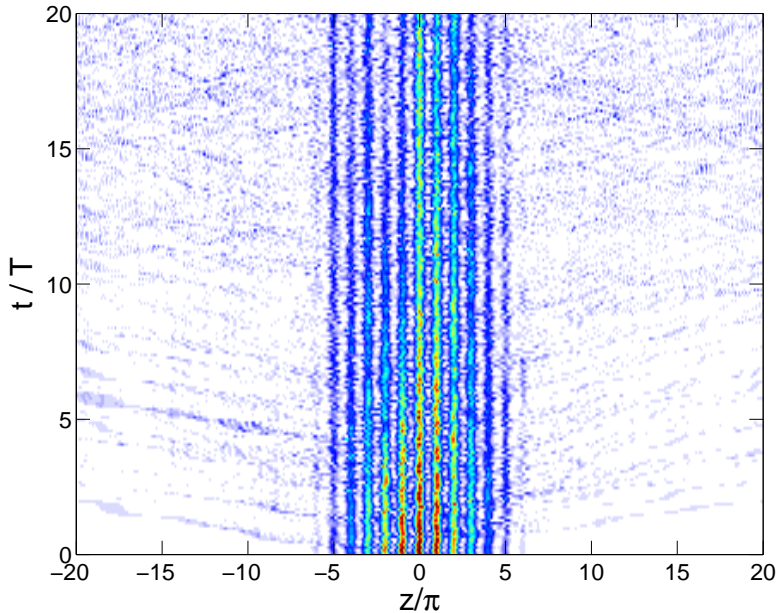}
\caption{Evolution of the same inital wave packet as in Fig.~\ref{fig:F_3}
	at the supposed second band collapse ($K_0 = 5.52$): Here the 
	multiphoton-like resonances visible in Fig.~\ref{fig:F_1} lead to 
	a marked degradation of the localization.}
\label{fig:F_5}
\end{figure}

It is then also of interest to monitor the evolution at the supposed second 
collapse, at $K_0 = 5.52$; this is done in Fig.~\ref{fig:F_5}. While the 
``regular spreading'' that has been prominent in Fig.~\ref{fig:F_3} indeed
seems to have stopped, small probability wavelets leak out of the initial 
packet almost immediately, spreading rapidly over the lattice. This is an 
effect of the multiphoton-like resonances previously spotted in 
Fig.~\ref{fig:F_1}, which assist parts of the wave function in getting to
higher bands, allowing them to escape on a short time scale. 

As long as interband transitions remain negligible, the resulting single-band 
dynamics can often be regarded as ``semiclassical''~\cite{AshcroftMermin76}: 
Namely, if an initial packet is strongly centered in $k$-space around some 
arbitrary wave number $\kc \equiv \kc(0)$, this center wave number evolves
in time according to Bloch's famous  ``acceleration theorem''
\begin{equation}
	\hbar \dot{k}_{\rm c}(t) = F(t) \; ,
\end{equation} 
similar to the evolution~(\ref{eq:IXH}) of the index of a single 
Houston state. The model Hamiltonian~(\ref{eq:HDL}) specifies 
$F(t) = F_1\cos(\omega t + \phi)$, so that in this case  
\begin{equation}
	\kc(t) = \kc(0) + \frac{F_1}{\hbar\omega}
	\Big( \sin(\omega t + \phi) - \sin(\phi) \Big) \; .
\label{eq:CLS}
\end{equation}
The packet's group velocity then is given by the derivative of the dispersion 
relation $E(k)$ of the band it lives in, evaluated at this moving center wave 
number~(\ref{eq:CLS}):
\begin{equation}
	v_{\rm group}(t) = \left. \frac{1}{\hbar} 
	\frac{\rd E}{\rd k} \right|_{\kc(t)} \; . 
\label{eq:SGV}	
\end{equation}
Taking the tight-binding relation~(\ref{eq:EDR}) as a good approximation for 
the actual lowest energy band of our model, this yields
\begin{equation}
	v_{\rm group}(t) = \frac{2Ja}{\hbar} \sin\!\Big( \kc(t) a \Big) \; .
\label{eq:OGV}
\end{equation} 
Upon time-averaging, one is therefore left with
\begin{eqnarray}
	\overline{v}_{\rm group} =
	\frac{2J_{\rm eff}a}{\hbar} 
	\sin\!\Big( \kc(0)a - K_0\sin(\phi) \Big) \; ,
\label{eq:AGV}
\end{eqnarray}	
where $J_{\rm eff}$ again is the driving-dependent effective hopping matrix
element~(\ref{eq:MHE}), and $K_0$ is the scaled amplitude~(\ref{eq:SCA}).
Thus, the initial phase $\phi$ may be utilized for imparting some momentum 
to the packet. Nonetheless, for any combination of $\kc(0)$ and $\phi$ the
average group velocity vanishes when $J_{\rm eff} = 0$, as corresponding to 
ideal dynamic localization.

\begin{figure}[ht!]
\centering
\includegraphics[width=0.63\textwidth]{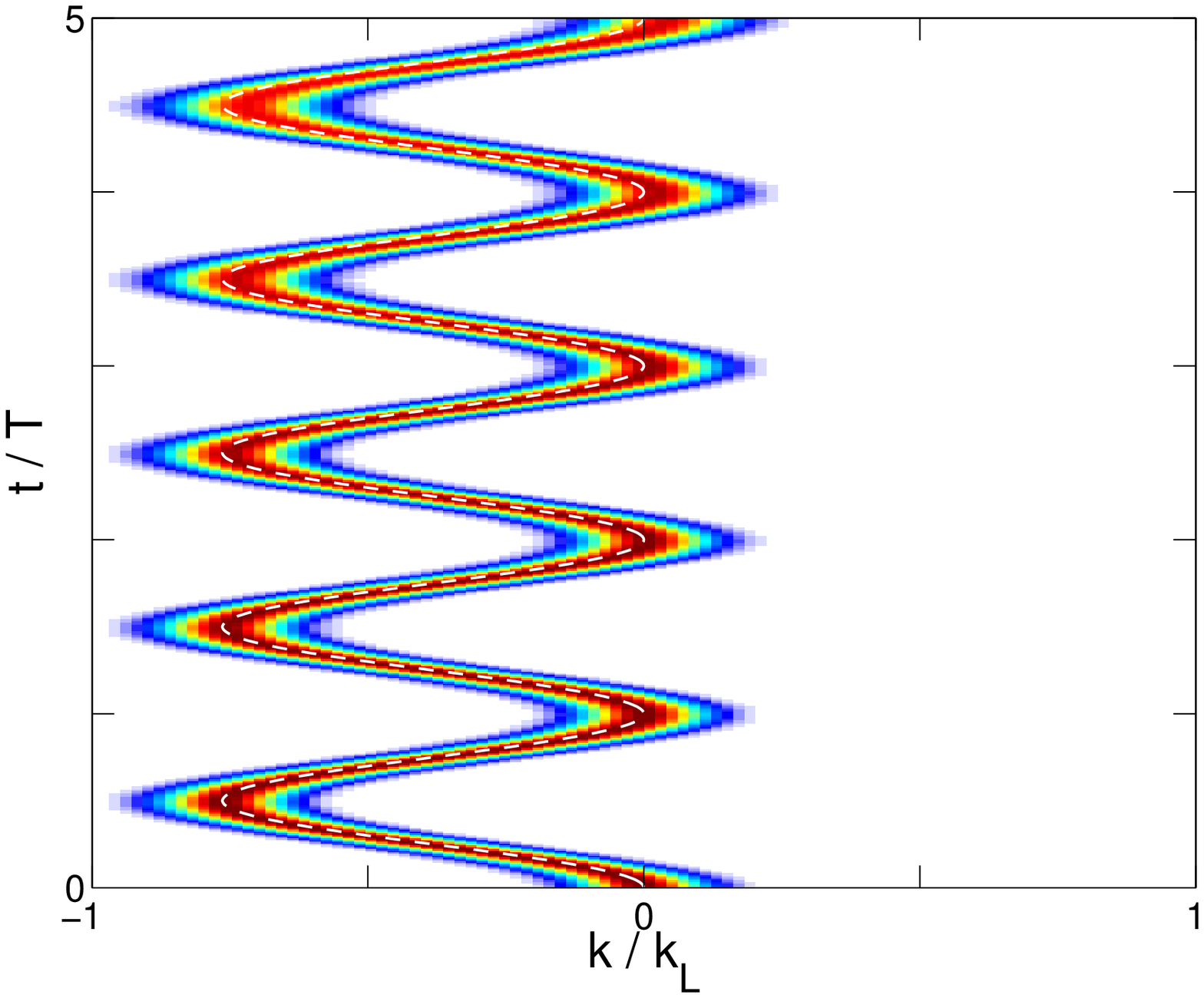}
\caption{Evolution of the $k$-space distribution initially given by
	Eq.~(\ref{eq:GKD}) with width $\Delta k /\kL = 0.1$, for $K_0 = 1.2$ 
	and $\kc/\kL = 0$. The white-dashed line indicates the ``classical'' 
	solution~(\ref{eq:CLS}).} 
\label{fig:F_6}
%\end{figure}

\strut

%\begin{figure}[b]
\centering
\includegraphics[width=0.63\textwidth]{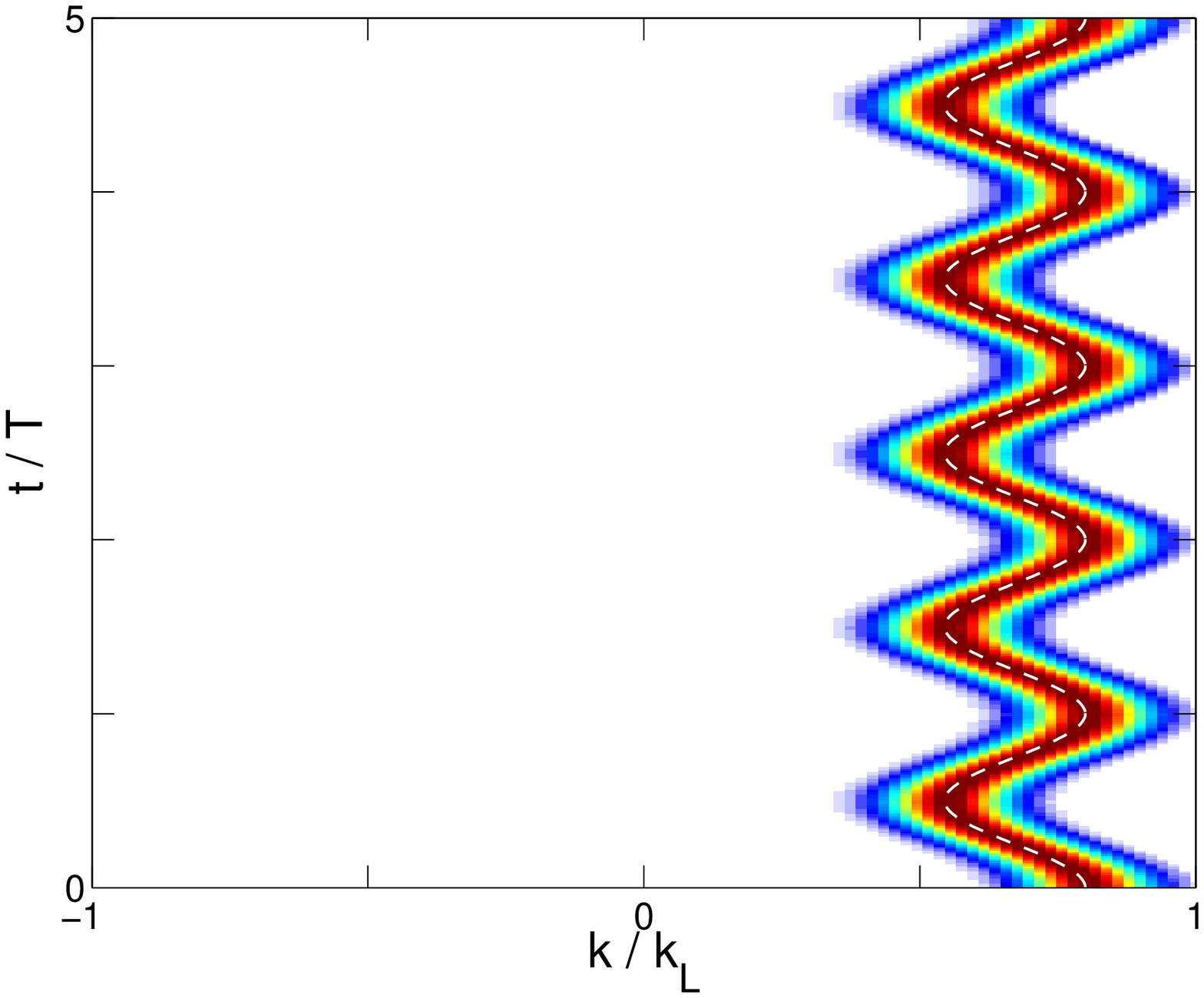}
\caption{Evolution of the $k$-space distribution initially given by
	Eq.~(\ref{eq:GKD}) with width $\Delta k /\kL = 0.1$, but now for 
	$K_0 = 0.4$ and $\kc/\kL = 0.8$, so that the white-dashed ``classical''
	solution~(\ref{eq:CLS}) starts from a non-zero value.}
\label{fig:F_7}
\end{figure}

\newpage

\begin{figure}[ht!]
\centering
\includegraphics[width=0.63\textwidth]{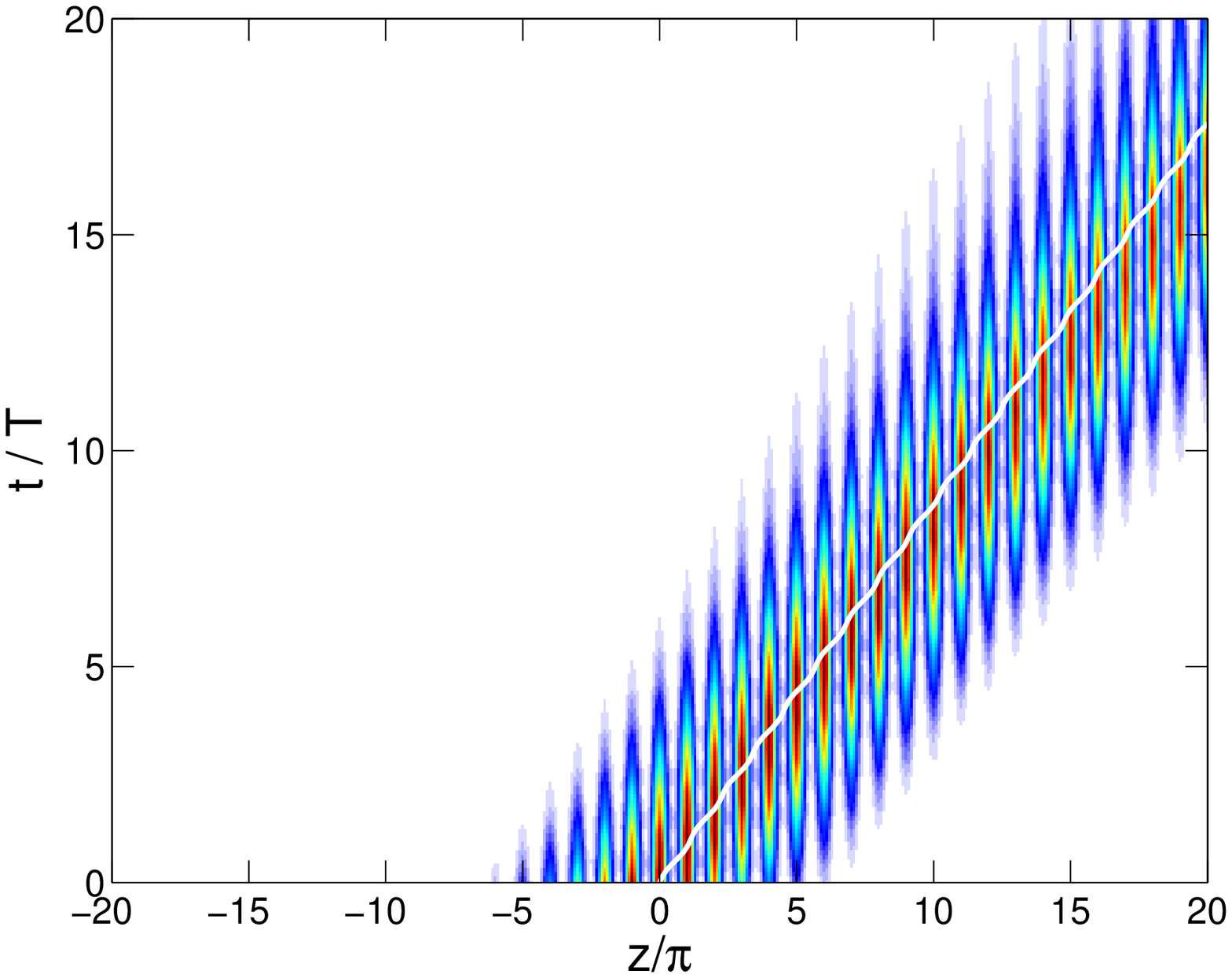}
\caption{Evolution of the Bloch wave packet~(\ref{eq:BWP}) with initial 
	$k$-space width $\Delta k/ \kL = 0.1$ and initial momentum 
	$\kc/\kL = 0.8$, driven with scaled amplitude $K_0 = 0.4$, 
	as corresponding to the $k$-space distribution depicted in 
	Fig.~\ref{fig:F_7}. The white line marks the trajectory obtained 
	by integrating the oscillating group velocity~(\ref{eq:OGV}).} 
\label{fig:F_8}
%\end{figure}

\strut

%\begin{figure}[b]
\centering
\includegraphics[width=0.63\textwidth]{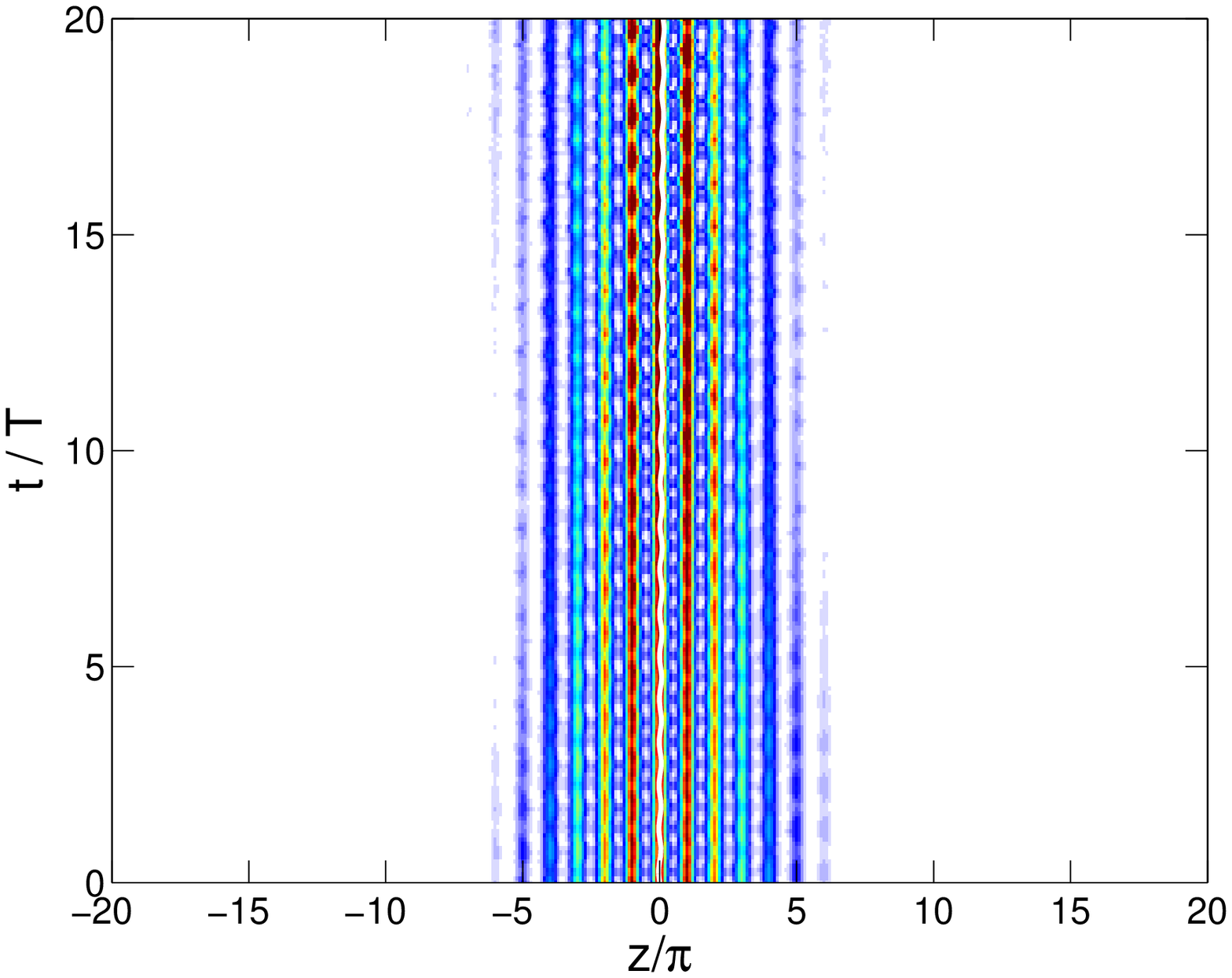}
\caption{Evolution of the Bloch wave packet~(\ref{eq:BWP}) with initial 
	$k$-space width $\Delta k/ \kL = 0.1$ and initial momentum 
	$\kc/\kL = 0.8$, now driven with scaled amplitude $K_0 = 2.35$: 
	Despite the nonzero average momentum, the average group velocity 
	vanishes because of the band collapse. The white line in the center 
	is obtained as described in Fig.~\ref{fig:F_8}.}
\label{fig:F_9}
\end{figure}

This semiclassical behavior is illustrated by a further set of figures. 
In Fig.~\ref{fig:F_6} we plot the evolution of the exact $k$-space density 
that originates from the initial condition~(\ref{eq:GKD}). Again we set  
$\Delta k /\kL = 0.1$, meaning that the distribution is sufficiently narrow
to ensure the validity of Eq.~(\ref{eq:SGV}); moreover, $\kc/\kL = 0$ and 
$K_0 = 1.2$. Since $\phi = \pi/2$, the distribution then oscillates around 
$\overline{k} = -F_1/(\hbar\omega)$, or $\overline{k}/\kL = -K_0/\pi$, 
following precisely the $k$-space trajectory predicted by Eq.~(\ref{eq:CLS}).

A nonzero average momentum of the packet can likewise be achieved by
selecting some suitable value of $\kc/\kL$. Figure~\ref{fig:F_7} shows an
example with $\kc/\kL = 0.8$, while $K_0 = 0.4$ and $\Delta k /\kL = 0.1$.     
This obviously corresponds to a wave function $\langle x | \psi(t) \rangle$
which moves into the positive $x$-direction all the time; the density of
this wave function is displayed in Fig.~\ref{fig:F_8}. Here the white line
indicates the classical trajectory that results from integrating the group
velocity~(\ref{eq:OGV}); indeed, this trajectory describes the motion of
the packet's center quite well. When adjusting the driving amplitude to
the first collapse, as in Fig.~\ref{fig:F_9}, the average motion stops 
despite the nonzero average momentum, as it should; when increasing $K_0$
to still higher values, so that $J_{\rm eff}$ becomes negative, the
packet's direction of motion can even be reversed.   

\begin{figure}[ht!]
\centering
\includegraphics[width=0.63\textwidth]{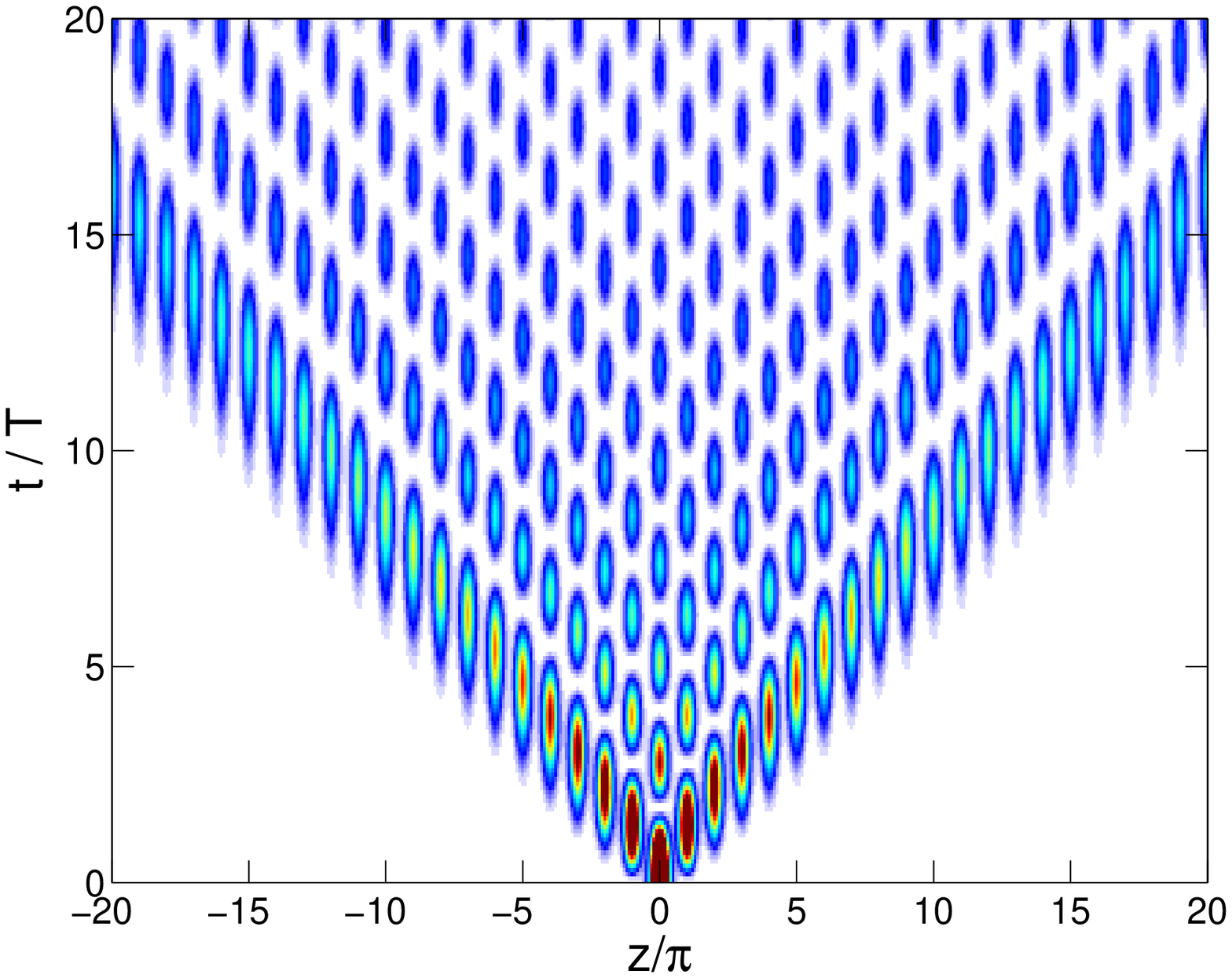}
\caption{Evolution of the wave function that originates from an initial
	single Wannier state in the undriven lattice. This state does not 
	possess a slowly varying envelope, and thus does not conform to 
	semiclassical dynamics.}
\label{fig:F_10}
%\end{figure}

\strut

%\begin{figure}[b]
\centering
\includegraphics[width=0.63\textwidth]{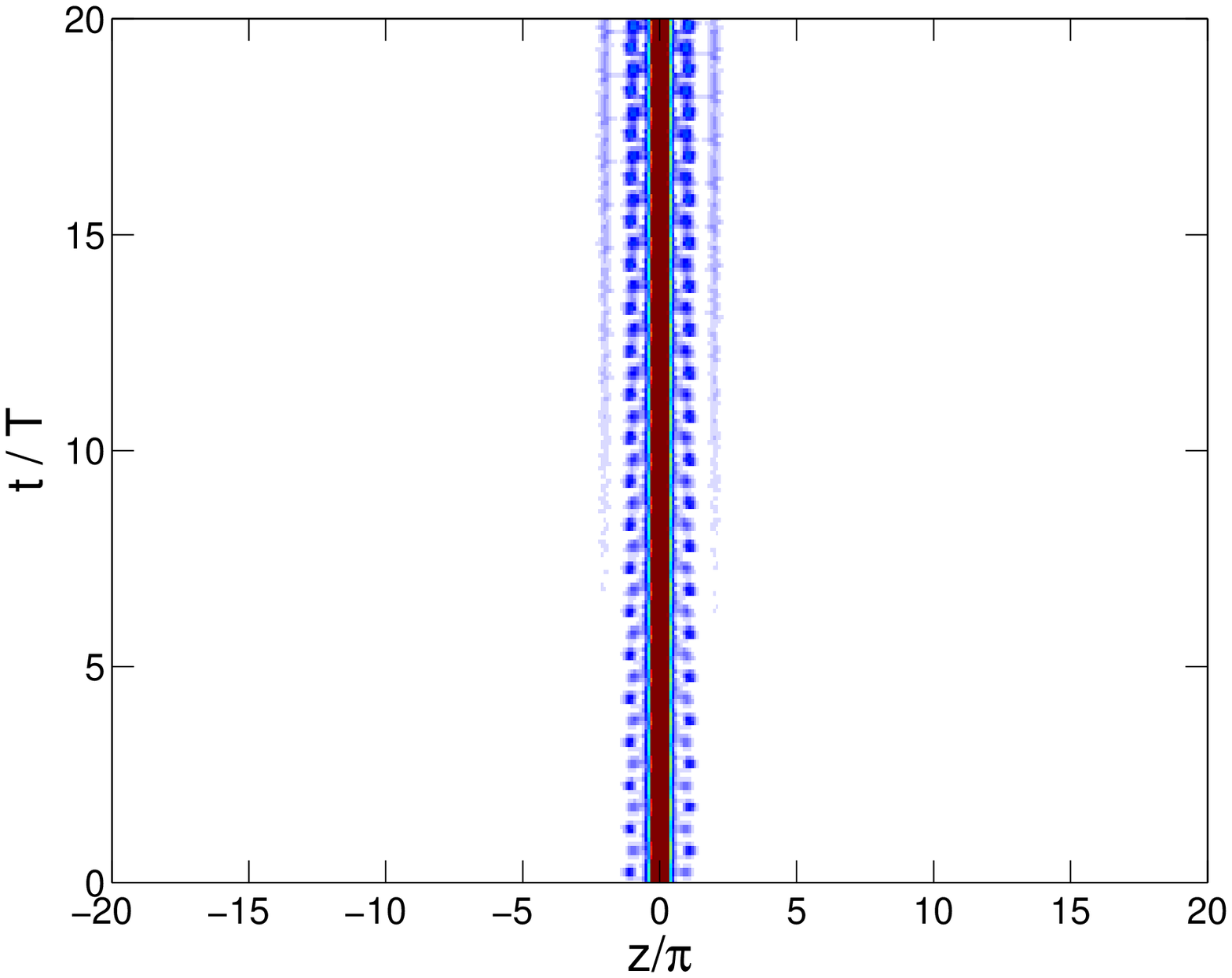}
\caption{Evolution of the wave function that originates from an initial
	single Wannier state when driven with scaled amplitude $K_0 = 2.35$.
	The semiclassical approximation is not applicable here, but
	dynamic localization works nonetheless.} 
\label{fig:F_11}
\end{figure}

While the semiclassical approach to dynamic localization may be helpful, 
insofar as it appeals to our intuition, its  explanation in terms of 
``prohibited dephasing'' resulting from a quasienergy band collapse is 
{\em much\/} more powerful: This view immediately reveals that not only 
does the average motion of a wave packet come to a complete standstill, but 
so does its spreading; moreover, prohibited dephasing applies to {\em any\/} 
initial condition, regardless whether or not its envelope varies suffiently 
slowly to justify the semiclassical approximation. As an extreme example 
of ``nonclassical'' motion we consider in Fig.~\ref{fig:F_10} the undriven
evolution of a wave function that coincides with a single Wannier function of 
the optical lattice~\cite{BoersEtAl07} at $t = 0$, and therefore certainly 
does not possess a slowly varying envelope then, giving rise to a fairly 
complex spreading pattern which differs substantially from the semiclassical 
one previously visualized in Fig.~\ref{fig:F_3}. Nonetheless, when driven 
with the amplitude $K_0 = 2.35$ marking the first quasienergy band collapse, 
one observes another occurrence of dynamic localization, as witnessed by 
Fig.~\ref{fig:F_11}; the difference between the two evolution patterns 
depicted in Figs.~\ref{fig:F_10} and~\ref{fig:F_11} could hardly be more 
striking.

\begin{figure}[t!]
\centering
\includegraphics[width=0.7\textwidth]{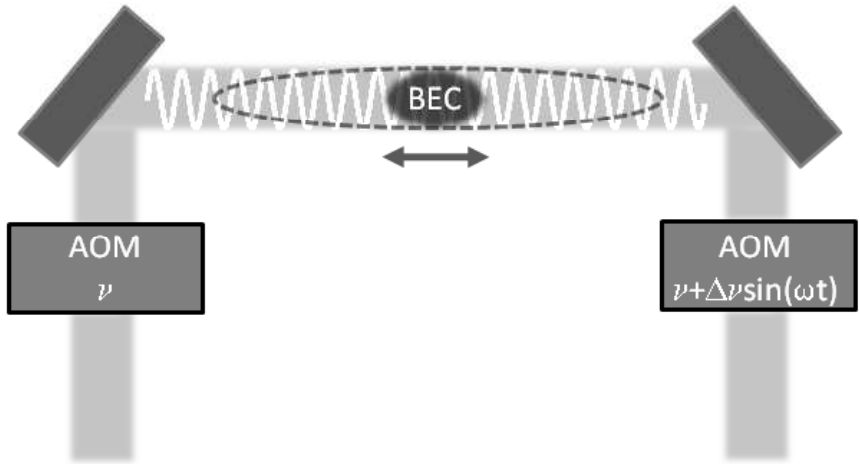}
\caption{Experimental setup for {\em in-situ\/} measurement of dynamic 
	localization of a Bose-Einstein condensate (BEC) in a driven optical 
	lattice: The frequencies of two laser beams are shifted with the 
	help of acousto-optic modulators (AOMs) by $\nu$ and by 
	$\nu + \Delta\nu\sin(\omega t)$, respectively, before being 
	directed against each other by mirrors. The resulting optical
	lattice then oscillates in the laboratory frame, giving rise to an
	oscillating inertial force in the frame of reference co-moving with
	the lattice. After the initial longitudinal confinement is switched
	off, the BEC expands by well-to-well tunneling; its final width
	(indicated by the dashed line) is recorded by imaging its shadow
	cast by a resonant flash onto a CCD chip. (Figure courtesy of 
	O.~Morsch.)}   
\label{fig:FExp_1}
\end{figure}

\begin{figure}[t!]
\centering
\includegraphics[width=1.0\textwidth]{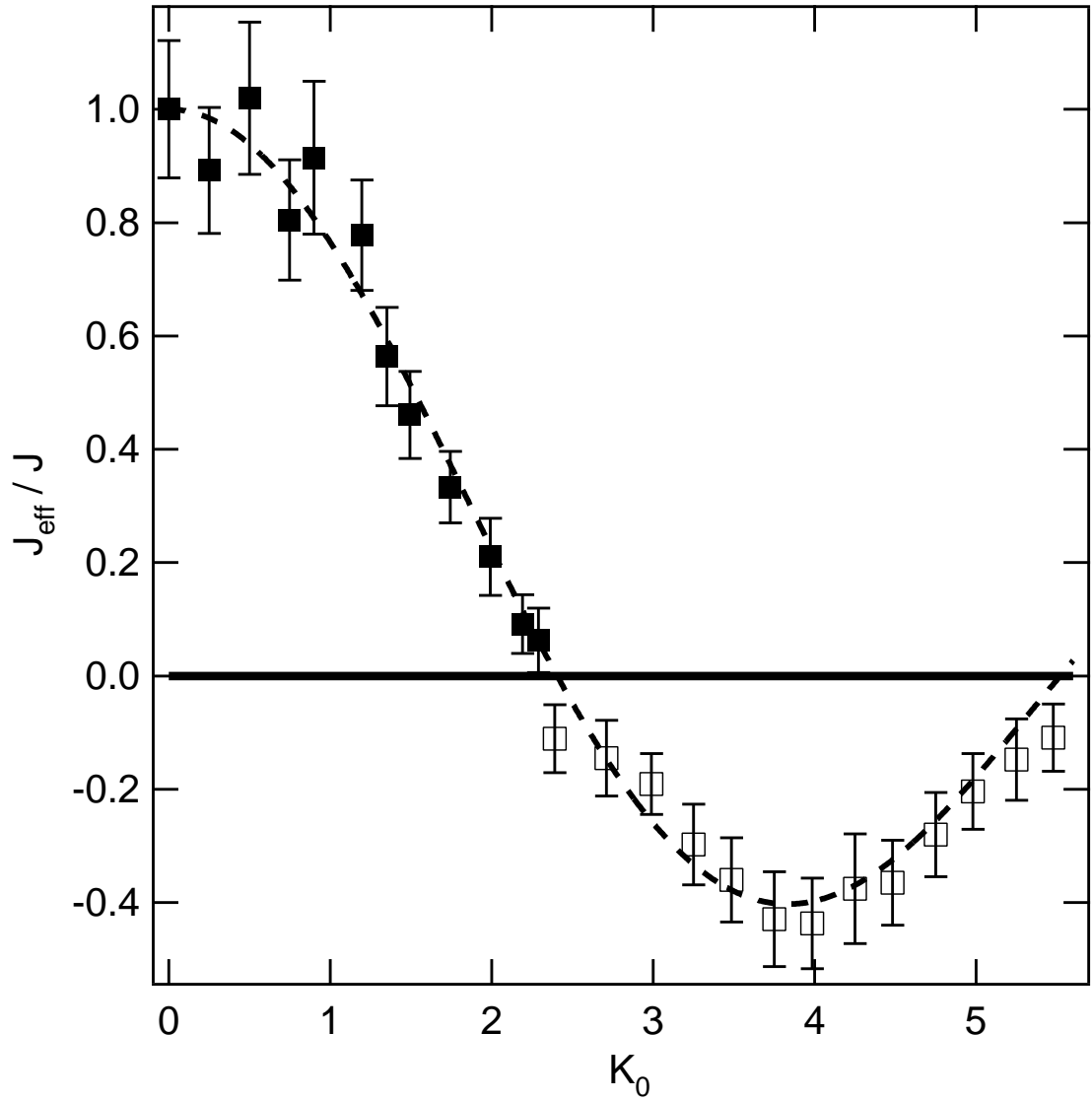}
\caption{Experimental results for the ratio $J_{\rm eff}/J$ of the effective 
	hopping matrix element~(\ref{eq:MHE}) to the bare one as a function 
	of the scaled driving amplitude~(\ref{eq:SCA}), obtained with a 
	Bose-Einstein condensate of $^{87}$Rb atoms in an optical lattice 
	with a depth of $V_0 = 6 \, \Er$ ($\lambda = 842$~nm), driven with 
	frequency  $\omega/(2\pi) = 4.0$~kHz. The dashed line corresponds 
	to the expected Bessel function ${\rm J}_0(K_0)$. (Figure courtesy 
	of O.~Morsch.)}
\label{fig:FExp_2}
\end{figure}

In actual laboratory experiments it is advantageous to work with a 
phase-coherent atomic Bose-Einstein condensate, rather than with individual 
atoms: If the density of the condensate is sufficiently low, or if the 
interatomic $s$-wave scattering length is tuned close to zero by means of a 
Feshbach resonance~\cite{RoatiEtAl08}, the condensate is practically ideal, so 
that one effectively can perform a measurement on an ensemble of identically
prepared noninteracting atoms in a single shot. Figure~\ref{fig:FExp_1} shows 
a possible experimental setup~\cite{LignierEtAl07,EckardtEtAl09}: The optical 
lattice is formed by two laser beams of wavelength~$\lambda$, which are 
directed against each other with the help of mirrors. Each beam passes through 
an acousto-optic modulator which shifts its frequency by $\nu$ and by
$\nu + \Delta \nu(t)$, respectively. Because of the frequency difference 
$\Delta\nu(t)$ thus introduced between the counterpropagating beams, the 
condensate experiences the potential
\begin{equation}
	V_{\rm lab}(x,t) = \frac{V_0}{2}\cos\left( 2\kL 
	\left[ 
	x + \frac{\lambda}{2}\int_0^t \! \rd \tau \, \Delta \nu(\tau)
	\right] \right)
\end{equation}
in the laboratory frame, which means that the lattice position shifts in time 
according to the prescribed protocol $\Delta\nu(t)$. In a frame of reference 
co-moving with the lattice, this shift translates into the inertial force
\begin{equation}
	F(t) = M \, \frac{\lambda}{2} \, \frac{\rd \Delta\nu(t)}{\rd t} \; .
\end{equation}
Therefore, choosing 
$\Delta \nu(t) = \Delta\nu_{\rm max} \sin(\omega t + \phi)$ leads to the
desired Hamiltonian~(\ref{eq:HDL}) in the co-moving frame, with the driving
amplitude
\begin{equation}
	F_1 = M\omega\,\frac{\lambda}{2}\,\Delta\nu_{\rm max} \; .
\end{equation}	
Now a Bose-Einstein condensate initially trapped in the center of the
oscillating lattice is allowed to expand freely in the lattice direction by 
well-to-well tunneling after switching off the longitudinal confinement, while 
maintaining a weak transversal confinement in order to keep the condensate
in the lattice. After a variable expansion time, the {\em in situ\/} width
of the condensate is determined by a resonant flash, the shadow cast by which 
is imaged onto a CCD chip~\cite{LignierEtAl07}. The measured expansion rate
then is to a good approximation proportional to $|J_{\rm eff}|$, that is,
to the absolute value of the effective hopping matrix element~(\ref{eq:MHE});
in principle, even the sign of $J_{\rm eff}$ can be deduced from additional 
time-of-flight measurements~\cite{LignierEtAl07}. In  Fig.~\ref{fig:FExp_2} 
we display data for the ratio $J_{\rm eff}/J$ acquired in this manner by the 
Pisa group with a condensate of $^{87}$Rb atoms in a lattice of depth 
$V_0/\Er = 6.0$, driven with frequency $\omega/(2\pi) = 4.0$~kHz, after 
expansion times of $150$~milliseconds. Evidently these data match the expected 
Bessel function ${\rm J}_0(K_0)$ quite well even up to the second zero. Note 
that here one has $\hbar\omega/\Er = 1.24$, so that the frequency employed in 
these measurements is significantly higher than in our model calculations. 
This means that the inequality $4J < \hbar\omega$ is satisfied in a 
stronger manner, while $\hbar\omega$ still remains reasonably small compared 
to the band gap. As a consequence, even the second band collapse can be quite 
well developed. In any case, this figure strikingly demonstrates that the 
concept of dynamic localization by now has crossed, in the context of 
mesoscopic matter waves, the threshold from an idealized theoretical concept 
to a well-controllable laboratory reality.

\clearpage 	
\section{What is it good for?}

Up to this point we have considered no more than a possible realization 
of dynamic localization which comes fairly close to the theoretical 
ideal~\cite{DunlapKenkre86}. Apart from its observation with dilute 
Bose-Einstein condensates in time-periodically shifted optical 
lattices~\cite{LignierEtAl07,EckardtEtAl09}, this type of quantum wave 
propagation has meanwhile also been made visible by means of an optical 
analog based on sinusoidally-curved lithium-niobate waveguide 
arrays~\cite{LonghiEtAl06}. This is certainly interesting, but it is not what 
one would call ``deep''; the ``prohibited dephasing''-view clearly reveals that
the only physics entering here is summarized by stating that an initial state 
is ``frozen'' in time if the phase factors of all of its spectral components 
evolve at the same speed. Yet, the accompanying band collapse furnishes a 
strong hint that there may be more in stock. Namely, when the ideal dynamics 
is somehow perturbed it is the bandwidth which sets the scale with respect 
to which the strength of such a perturbation has to be gauged. A prominent 
example is provided by the repulsive interaction between ultracold atoms in 
an optical lattice; the strength of this interaction is expressed in terms 
of a parameter~$U$ which quantifies the repulsion energy of one pair of 
atoms occupying the same lattice site~\cite{BlochEtAl08}. 
Accordingly, the characteristic dimensionless parameter then is the ratio 
$U/J$; here $J = W/4$ is taken instead of the bandwidth~$W$. Indeed, it is 
this ratio $U/J$ which decides which quantum phase a gas of ultracold, 
repulsively interacting atoms in an optical lattice adopts: For $U/J \ll 1$ 
the system is superfluid, but becomes a Mott insulator when this ratio exceeds 
a critical value~\cite{BlochEtAl08}. Hence, when recalling that $J$ is 
replaced by the effective hopping matrix element~(\ref{eq:MHE}) when the 
system is driven with appropriate parameters, it is only natural to predict 
that this superfluid-to-Mott insulator transition can be induced in a 
time-periodically shifted optical lattice by varying the driving 
force~\cite{EckardtEtAl05,CreffieldMonteiro06}: Assuming that one 
starts in the superfluid phase, $J_{\rm eff}$ can then virtually be made 
arbitrarily small by adjusting the scaled amplitude $K_0$ to a zero of 
${\rm J}_0$, resulting in a value of $U/J_{\rm eff}$ so large that the system 
is forced to enter the Mott regime. The experimental confirmation of this 
scenario, achieved by the Pisa group~\cite{ZenesiniEtAl09}, probably 
constitutes the first known example of coherent control exerted by means of 
time-periodic forcing on a quantum phase transition. 
    
There are other types of perturbations, associated with deviations from
perfect translational symmetry, which affect even noninteracting ultracold 
atoms in optical lattices. Most notably, the system governed by the 
tight-binding Hamiltonian
\begin{equation}
	H_{\rm AA} = -J \sum_{\ell} \Big( |\ell + 1 \rangle \langle \ell |
	+ | \ell \rangle \langle \ell + 1 | \Big) 
	+ V \sum_{\ell} \cos(2\pi g\ell + \delta) 
	| \ell \rangle \langle \ell | \; ,
\label{eq:HAA}
\end{equation}
differing from its antecedent~(\ref{eq:TBS}) through additional on-site 
energies which oscillate along the lattice with amplitude~$V$, shows a quite 
peculiar behavior when the number~$g$ is irrational, so that this system 
becomes {\em quasiperiodic\/}~\cite{Harper55,AubryAndre80,Sokoloff85}: As 
long as $|V/J| < 2$, so that the on-site perturbations are relatively weak, 
all of its energy eigenstates still remain extended over the entire lattice 
in a Bloch-like manner, whereas they are all exponentially localized, 
with one common localization length, when $|V/J| > 2$. Thus, there is a 
metal-insulator-like, incommensurability-induced transition at $|V/J| = 2$, 
originally studied by Harper~\cite{Harper55} in the context of conduction
electrons in a magnetic field, and later by Aubry and 
Andr\'{e}~\cite{AubryAndre80}; this transition can be realized approximately 
with ultracold atoms in a {\em bichromatic\/} optical lattice described by 
the potential 
\begin{equation}
	V_{\rm bic}(x) = \frac{V_0}{2}\cos(2\kL x)
	+ V_1 \cos(2g \kL x + \delta) \; .
\label{eq:BIP}
\end{equation}
The guiding idea here is to employ a primary lattice with depth~$V_0$
for setting up the hosting tight-binding system~(\ref{eq:TBS}), as before, 
and then to invoke a secondary lattice with much smaller depth $2 V_1$ for 
achieving the required modulation of the local energies at the sites of the
host~\cite{DreseHolthaus97a,DreseHolthaus97b}. When the primary lattice is 
comparatively shallow, possessing a depth of only a few recoil energies, 
the transition occurs stepwise upon increasing $V_1$~\cite{BoersEtAl07}, 
featuring pronounced mobility edges resulting mainly from the next-to-nearest
neighbor couplings between the host's sites which are present in the full 
bichromatic potential~(\ref{eq:BIP}), but do not occur in the Aubry-Andr\'{e} 
model~(\ref{eq:HAA}). When $V_0/\Er \gg 1$, so that the primary lattice 
is so deep that these additional couplings may be safely neglected, the
transition occurring in the actual bichromatic lattice~(\ref{eq:BIP})
is fairly sharp. The parameter~$J$ then again is given approximately by 
Eq.~(\ref{eq:MAP}); moreover, one has 
\begin{equation}
	V/\Er \sim \frac{V_1}{\Er} 
	\exp\left(- \frac{g^2}{\sqrt{V_0/\Er}}\right) 
	\qquad \mbox{for} \quad V_0/\Er \gg 1 
\end{equation}
with reasonably chosen~$g$ on the order of unity. Therefore, the equation
$|V/J| = 2$ marking the metal-insulator-like transition in the ideal
Aubry-Andr\'{e} model now translates into the estimate~\cite{BoersEtAl07}
\begin{equation}
	\frac{V_1^{\rm c}}{\Er} \sim \frac{8}{\sqrt{\pi}}
	\left(\frac{V_0}{\Er}\right)^{3/4}
	\exp\!\left(- 2\sqrt{\frac{V_0}{\Er}} 
	            + \frac{g^2}{\sqrt{V_0/\Er}} \right) 
\label{eq:EST}
\end{equation}
for the critical strength~$V_1^{\rm c}$ of the secondary optical lattice, 
given a sufficient depth of the primary one. Indeed, this transiton has been 
observed with a Bose-Einstein condensate consisting of $^{39}$K atoms, using 
a magnetically tunable Fesbach resonance for rendering these atoms practically
noninteracting~\cite{RoatiEtAl08}.

When ultracold atoms in such a bichromatic lattice~(\ref{eq:BIP}) are 
subjected to time-periodic forcing, one obtains an additional knob which can 
be turned to induce the transition: Because~$J$ is replaced by the effective 
hopping strength~(\ref{eq:MHE}) when the system is suitably driven, one 
can cross the critical border $|V/J_{\rm eff}| = 2$ by varying the parameters 
of the driving force; the critical parameters then are linked approximately 
by the relation
\begin{equation}
	| {\rm J}_0(K_0) | \; \sim \; \frac{\sqrt{\pi}}{8} \frac{V_1}{\Er} 
	\left( \frac{V_0}{\Er} \right)^{-3/4}
	\exp\!\left(+ 2\sqrt{\frac{V_0}{\Er}} 
	            - \frac{g^2}{\sqrt{V_0/\Er}} \right) \; . 
\label{eq:REL}
\end{equation}
Hence, it is feasible to coherently control the metal-insulator-like 
transition exhibited by noninteracting ultracold atoms in properly designed 
bichromatic optical potentials through time-periodic 
forcing~\cite{DreseHolthaus97a,DreseHolthaus97b}. In order to substantiate 
this prediction, we now display the results of further numerical wave-packet 
calculations. In all of these we employ a primary lattice with depth 
$V_0/\Er = 5.7$, as in our preceding studies, and fix the incommensurability 
parameter at the golden mean $g = (\sqrt{5} - 1)/2$ up to numerical accuracy. 
With this choice, the above estimate~(\ref{eq:EST}) yields 
$V_1^{\rm c}/\Er \approx 0.165$ for the critical strength of the secondary 
lattice. The driving frequency is given by $\hbar\omega/\Er = 0.5$ throughout.

\begin{figure}[ht!]
\centering
\includegraphics[width=0.63\textwidth]{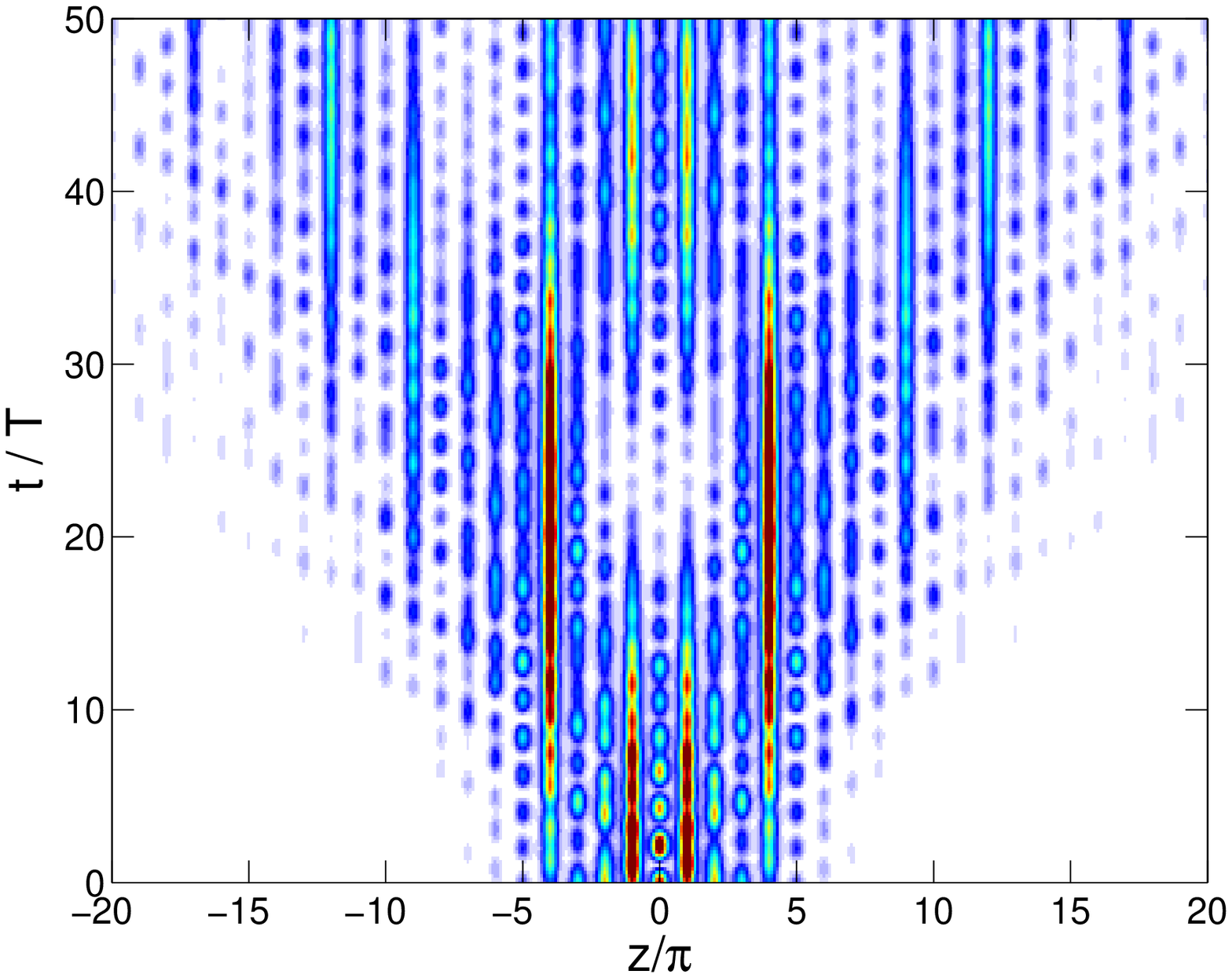}
\caption{Evolution of the same initial wave packet as in Fig.~\ref{fig:F_3}
	in an undriven bichromatic optical lattice~(\ref{eq:BIP}). Here
	the strength of the secondary potential is $V_1/\Er = 0.10$,
	so that the system is in its mobile ``metallic'' phase, allowing
	the wave function to spread.}
\label{fig:F_12}
%\end{figure}

\strut

%\begin{figure}[b]
\centering
\includegraphics[width=0.63\textwidth]{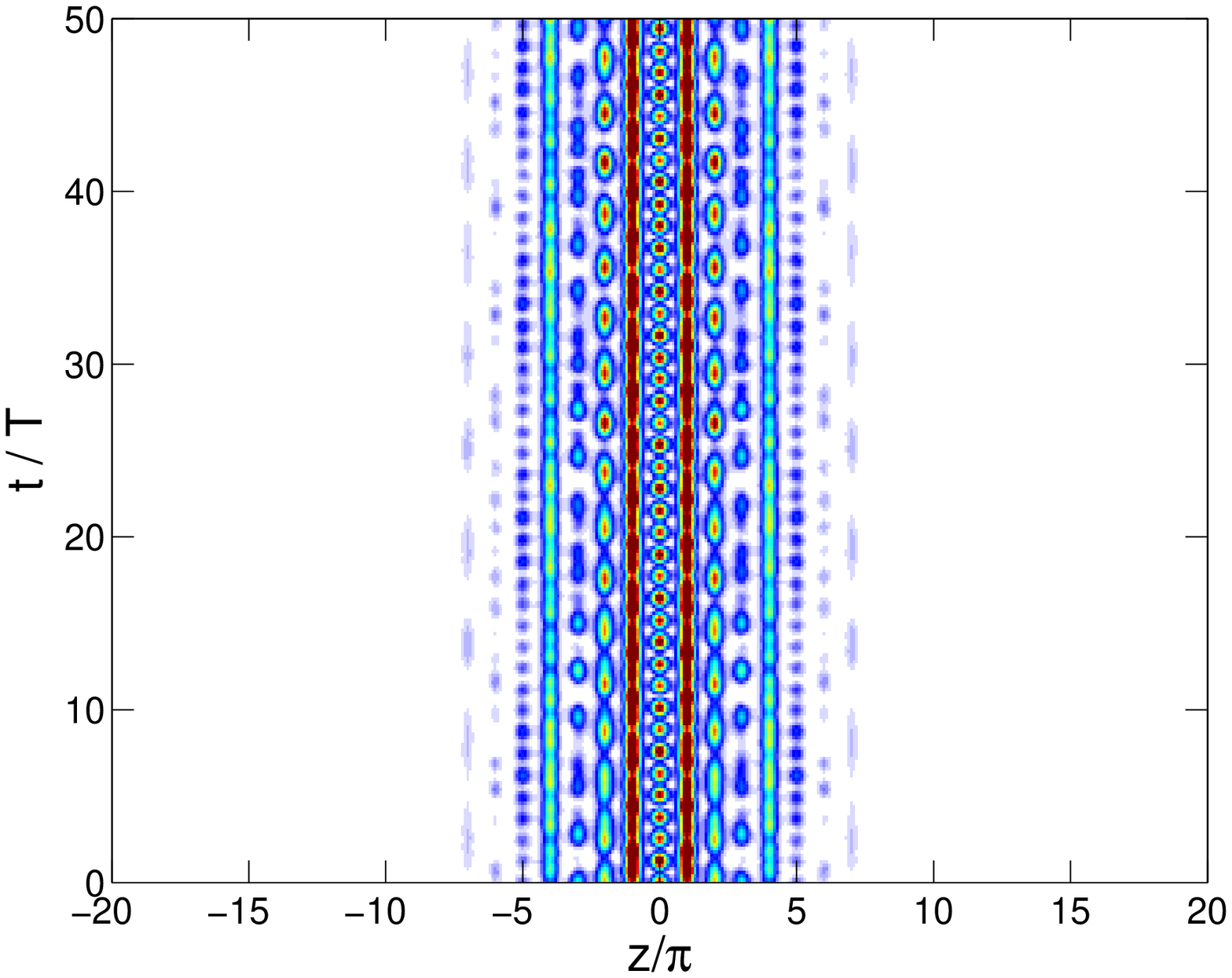}
\caption{Evolution of the same initial wave packet as in Fig.~\ref{fig:F_3}
	in an undriven bichromatic optical lattice~(\ref{eq:BIP}). Here
	the strength of the secondary potential is $V_1/\Er = 0.25$,
	so that the system is in its ``insulating'' phase, keeping the
	wave function localized.}
\label{fig:F_13}
\end{figure}

\begin{figure}[t!]
\centering
\includegraphics[width=0.63\textwidth]{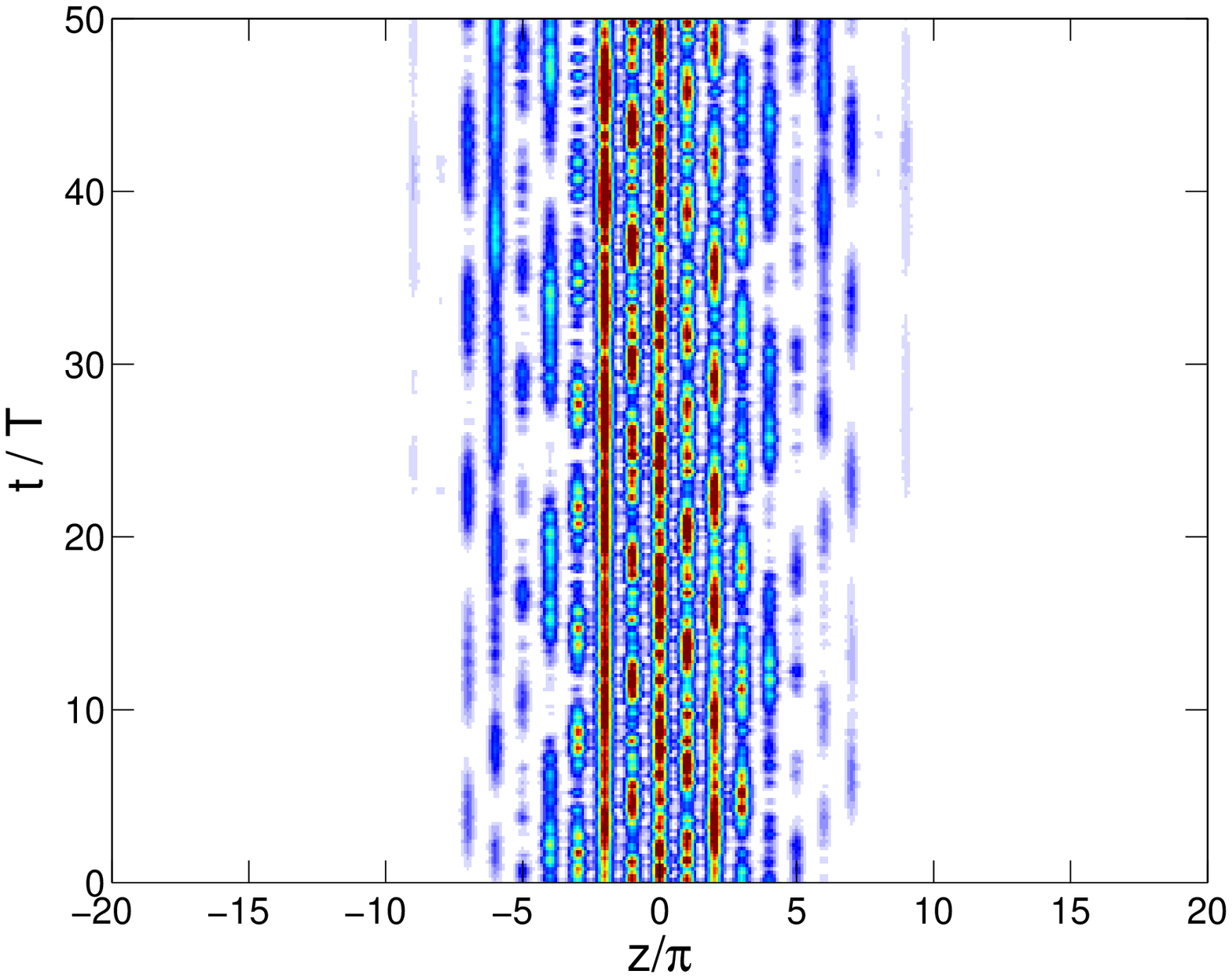}
\caption{Evolution of the same initial wave packet as in Fig.~\ref{fig:F_3}
	in a bichromatic optical lattice~(\ref{eq:BIP}) driven with
	scaled amplitude $K_0 = 1.7$. The strength of the secondary
	lattice is $V_1/\Er = 0.10$, as in Fig.~\ref{fig:F_12}, so 
	that the system would be in its ``metallic'' phase if there
	were no forcing.}  	
\label{fig:F_14}
\end{figure}

Figure~\ref{fig:F_12} visualizes the evolution of a wave function that
originates from the same Gaussian initial state as already employed in 
Fig.~\ref{fig:F_3}. Here the driving force is still absent, and the depth of 
the secondary lattice is $V_1/\Er = 0.10$, placing the system in its metallic 
phase; accordingly, the wave function readily explores the entire lattice. In 
contrast, when $V_1/\Er = 0.25$ and the drive is still switched off, the wave 
function remains localized as shown in Fig.~\ref{fig:F_13}; this indicates 
that we are encountering the insulating phase now. But the wave function 
also remains localized when the  secondary lattice is tuned back to 
$V_1/\Er = 0.10$ and the driving force acts with scaled amplitude $K_0 = 1.7$, 
as depicted in Fig.~\ref{fig:F_14}: The relation~(\ref{eq:REL}) predicts the 
transition from the metallic to the insulating phase to have occurred already 
at about $K_0 \approx 1.3$. It should be noted that there is a pronounced 
difference from the ideal dynamic localization reviewed in the preceding 
section: There the wave packet remains localized only when $K_0$ is exactly 
equal to a zero of ${\rm J}_0$. In contrast, here one switches from the
metallic into the insulating phase already when $|{\rm J}_0(K_0)|$ becomes 
sufficiently small. 
                      
\begin{figure}[ht!]
\centering
\includegraphics[width=0.63\textwidth]{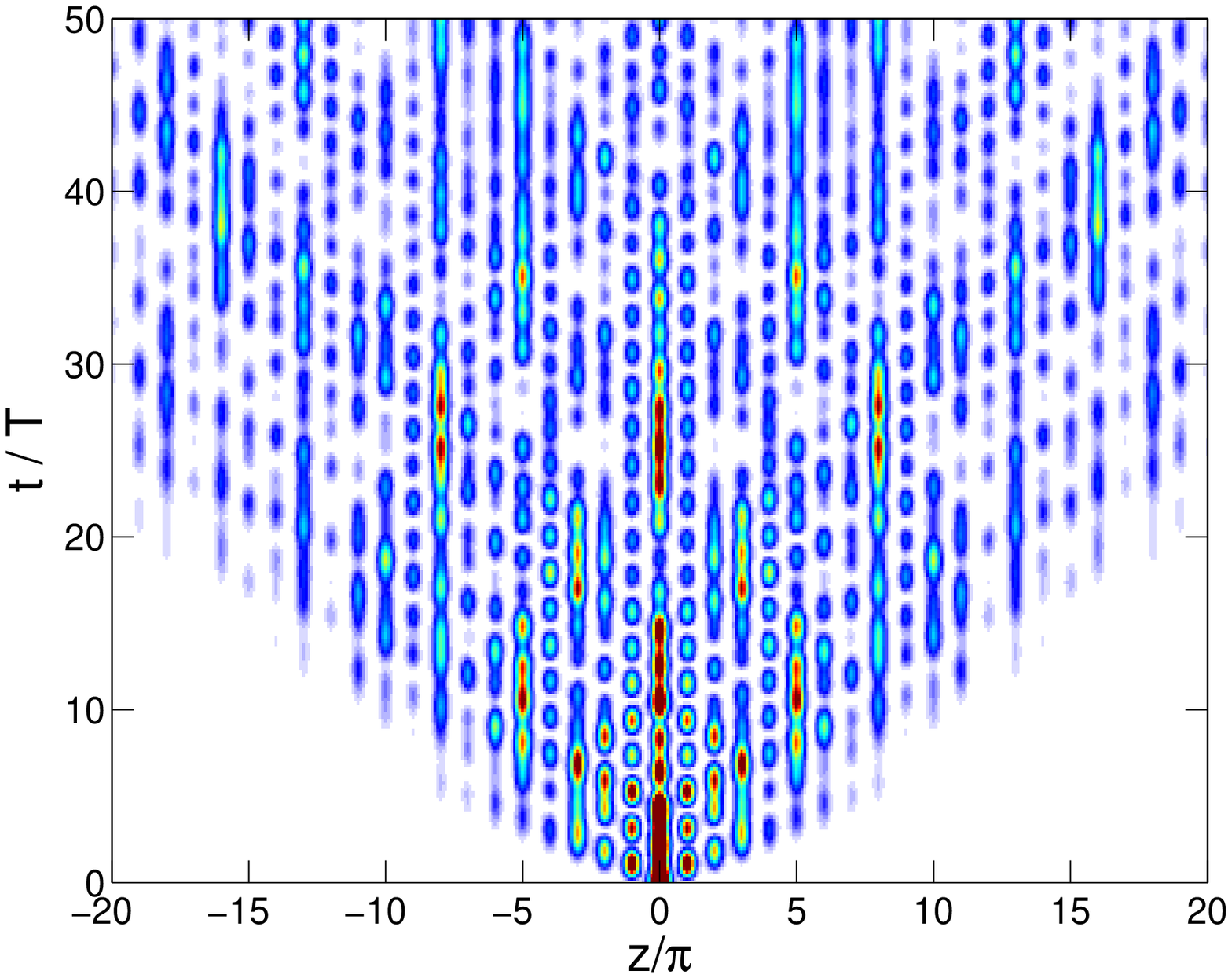}
\caption{Evolution of the wave function originating from a single Wannier
	state of the primary lattice in an undriven bichromatic optical 
	lattice~(\ref{eq:BIP}). Here the strength of the secondary potential 
	is $V_1/\Er = 0.10$, so that the system is in its ``metallic'' phase.}
\label{fig:F_15}
%\end{figure}

\strut

%\begin{figure}[b]
\centering
\includegraphics[width=0.63\textwidth]{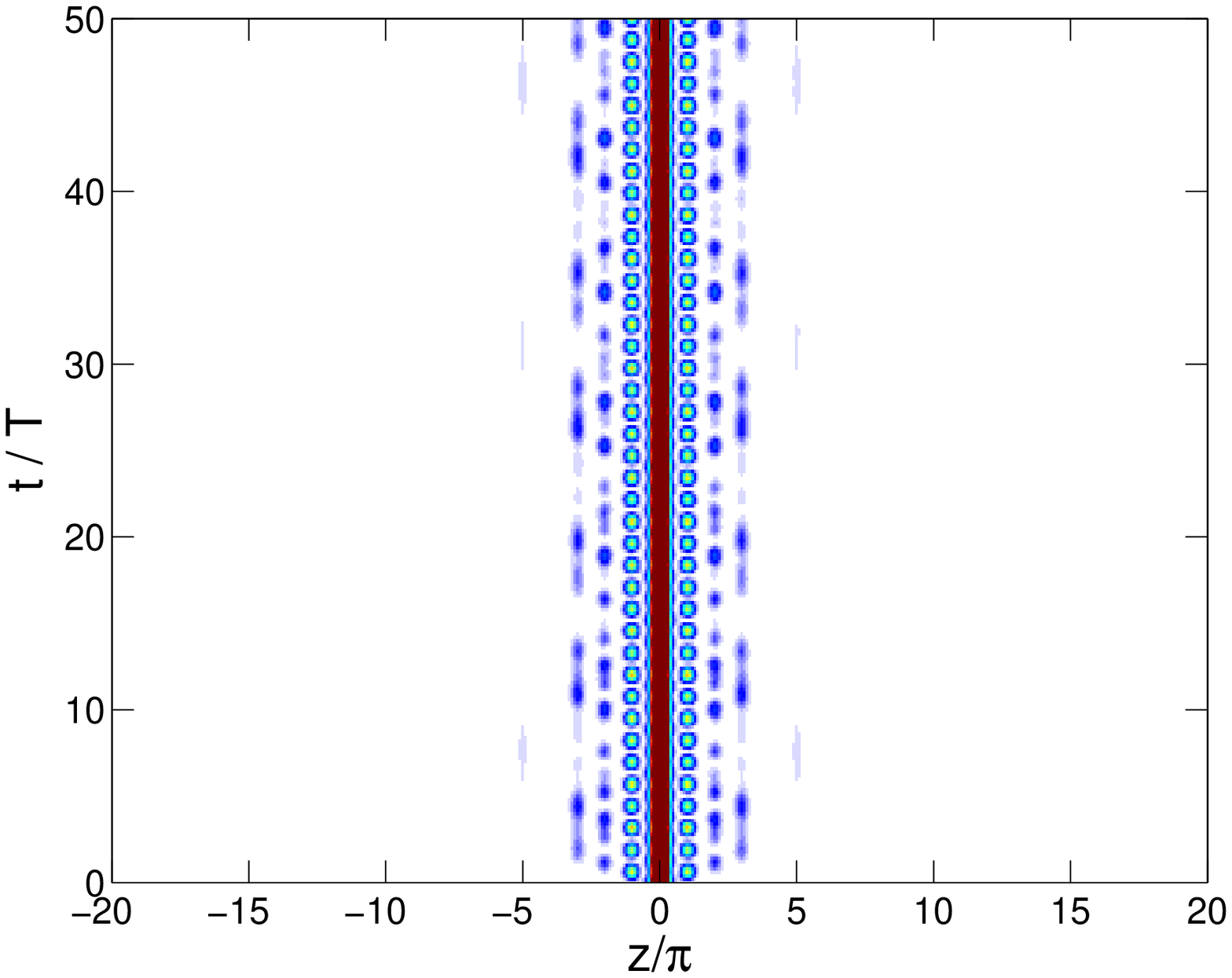}
\caption{Evolution of the wave function originating from a single Wannier 
	state of the primary lattice in an undriven bichromatic optical 
	lattice~(\ref{eq:BIP}). Here the strength of the secondary potential 
	is $V_1/\Er = 0.25$, so that the system is in its ``insulating'' 
	phase.}
\label{fig:F_16}
\end{figure}

\begin{figure}[t!]
\centering
\includegraphics[width=0.63\textwidth]{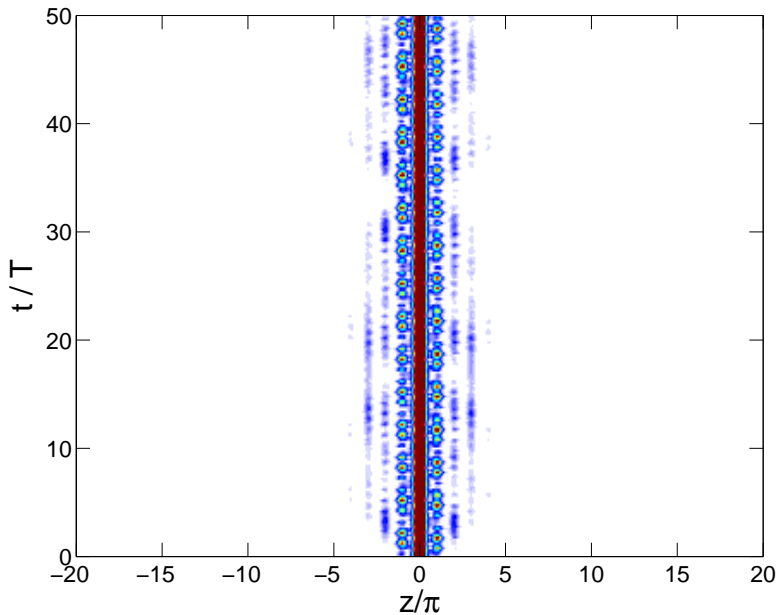}
\caption{Evolution of the wave function originating from a single Wannier 
	state of the primary lattice in a bichromatic optical 
	lattice~(\ref{eq:BIP}) driven with scaled amplitude $K_0 = 1.7$. 
	The strength of the secondary lattice is $V_1/\Er = 0.10$, as in 
	Fig.~\ref{fig:F_15}, so that the system would be in its ``metallic'' 
	phase if there were no forcing.}
\label{fig:F_17}
\end{figure}

Finally, we show a corresponding sequence of results for wave functions 
which evolve from an initial Wannier state of the primary lattice. In 
Fig.~\ref{fig:F_15} we again consider an undriven bichromatic lattice with 
$V_1/\Er = 0.10$, so that the mobile metallic phase enables uninhibited 
spreading; in Fig.~\ref{fig:F_16}, where $V_1/\Er = 0.25$, the system's 
insulating character then keeps the wave function strongly localized. But 
that same high degree of localization may also be obtained when again 
resetting the strength of the secondary lattice to $V_1/\Er = 0.10$, 
and switching on the driving force with scaled amplitude~$K_0 = 1.7$, 
as done in Fig.~\ref{fig:F_17}.

These figures vividly illustrate the main message: In the presence of 
time-periodic forcing it is the width of the underlying quasienergy band 
which determines the effective strength of deviations from perfect spatial 
periodicity. In an ideal lattice without such deviations one encounters 
``only'' dynamic localization, but in lattices with isolated, quasiperiodic, 
or random perturbations the strengths of these can be adjusted at will by 
suitably selecting the parameters of the drive. With regard to experimental 
tests, the enormous flexibility offered by ultracold atoms in optical 
potentials makes such systems far superior to electrons in ac-driven crystal 
lattices.   
 
When the concept of controlling the incommensurability-induced metal-insulator 
transition exhibited by the Aubry-Andr\'{e} model~(\ref{eq:HAA}) by means of 
time-periodic forcing was conceived~\cite{DreseHolthaus97a,DreseHolthaus97b} 
the experimental investigation of ultracold atoms in optical lattices was still
in its infancies. But now that this transition has been unambiguously observed 
with a non\-interacting Bose-Einstein condensate~\cite{RoatiEtAl08}, the 
demonstration of its coherent control has come into immediate reach. 
Besides the already established coherent control of the interaction-induced 
superfluid-to-Mott insulator transition~\cite{ZenesiniEtAl09}, this  
demonstration would constitute a further milestone achievement in the 
on-going effort to explore the newly emerging prospects provided by dressed 
matter waves.  

\vspace{8ex}

\noindent  
{\Large {\bf Acknowledgments}}

\noindent
We thank O.~Morsch and E.~Arimondo for continuing discussions of 
their experiments~\cite{LignierEtAl07,EckardtEtAl09,ZenesiniEtAl09}, 
and O.~Morsch in particular for providing Figs.~\ref{fig:FExp_1} 
and~\ref{fig:FExp_2}.   
This work was supported by the Deutsche Forschungsgemeinschaft under
Grant No.~HO~1771/6.

\end{doublespace}
%\bibliographystyle{plain}
%\bibliography{glmm,anders}
%\printindex

\end{document}